\documentclass[a4paper,fleqn]{cas-sc}

\usepackage[percent]{overpic}
\usepackage{caption}
\usepackage[authoryear,longnamesfirst]{natbib}
\usepackage{placeins}
\def\tsc#1{\csdef{#1}{\textsc{\lowercase{#1}}\xspace}}
\tsc{WGM}
\tsc{QE}
\tsc{EP}
\tsc{PMS}
\tsc{BEC}
\tsc{DE}

\begin{document}
\let\WriteBookmarks\relax
\def\floatpagepagefraction{1}
\def\textpagefraction{.001}

\shorttitle{Gender Bias in YouTube Exposure}

\shortauthors{Tan et~al.}

\title [mode = title]{Gender Bias in YouTube Exposure: Allocative and Structural Inequalities in Political Information Environments}

\tnotemark[1]

\tnotetext[1]{This document is the results of the research
   project funded by the General Project of Ministry of Education Foundation on Humanities and Social Sciences (23YJA860011), the Fundamental Research Funds for the Central Universities (1243200012), and the Guangdong Philosophy and Social Science Foundation Regular Project (GD24XXW02).}
%

\author[1,2]{Jipeng Tan}[style=chinese]


\ead{202431021021@mail.bnu.edu.cn}

\credit{Conceptualization, Data curation, Formal analysis, Methodology, Visualization, Writing – original draft, Writing – review and editing}

\affiliation[1]{
    organization={Computation Communication Research Center, Beijing Normal University},
    city={Zhuhai},
    postcode={519087},
    country={People's Republic of China}
}

\affiliation[2]{
    organization={School of Journalism and Communication, Beijing Normal University},
    city={Beijing},
    postcode={100875},
    country={People's Republic of China}
}

\affiliation[3]{
    organization={Faculty of  Arts and Sciences, Beijing Normal University},
    city={Zhuhai},
    postcode={519087},
    country={People's Republic of China}
}

\author[1,2]{Weifeng Zhang}[style=chinese,orcid=0009-0000-5170-8306]
\credit{Methodology, Supervision, Validation, Writing – review and editing}
\ead{WeifengZhang993@163.com}

\author[1,2]{Ye Wu}[style=chinese]
\credit{Conceptualization, Project administration, Resources}
\ead{wuye@bnu.edu.cn}

\author[3]{Jialin Guo}[style=chinese]
\credit{Data curation}
\ead{202211038397@mail.bnu.edu.cn}

\author[1,2]{Yong Min}[style=chinese,orcid=0000-0002-9387-3921]
\ead{myong@bnu.edu.cn}
\cortext[cor1]{Corresponding author}
\credit{Conceptualization, Funding acquisition, Investigation,  Project administration, Resources, Software, Supervision, Validation, Writing – review and editing}
\cormark[1]

\begin{abstract}
Recommendation algorithms have become the dominant mechanism for information distribution on digital platforms, profoundly shaping personalized information consumption environments. However, gender bias, as a significant form of algorithmic discrimination, may cause users to experience unequal exposure within different political information environments. Taking YouTube as a case, we conduct a controlled social-bot field experiment, where male-coded and female-coded profiles are constructed. We track the exposure and click patterns of these bots to analyze their recommendation trajectories. We analyze the distribution of recommended content from two dimensions: allocative bias and structural bias. First, we find statistically significant differences in allocative bias across male-coded and female-coded profiles, particularly in terms of issue distribution, ideological orientation, and political entities. Secondly, we observe structural bias in the political information environments, characterized by distinct clustering patterns. Additionally, time-series analysis shows that exposure pathways continue to be shaped over time by both communities detected in the co-occurrence network and individual profile-level dynamics. Finally, we construct a simple collaborative-filtering model that reproduces the observed gender bias. We argue that gender bias in recommendation systems is reflected not only in the allocation of political content, but also in how community structures shape these environments, reinforcing societal inequalities and highlighting the need for algorithmic fairness.

\end{abstract}



\begin{keywords}
YouTube \sep Algorithmic Audits \sep Recommendation systems \sep Gender Bias
\end{keywords}

\maketitle

\section{Introduction}

Recommendation systems have become a critical infrastructure for organizing information exposure in platformized communication environments \citep{RN13,RN17}. Unlike the relatively stable distribution logic of traditional media, platforms continuously track user behavior, dynamically adjust content rankings, and update subsequent distribution outcomes through repeated interactions, thereby profoundly shaping how the public encounters news and public affairs information \citep{RN3,RN44}. Recommendation systems therefore should not be understood merely as technical tools for improving information-matching efficiency; rather, they should be seen as key mechanisms for allocating information visibility, organizing exposure pathways, and shaping information environments \citep{RN59}. This is especially important in the context of political communication. On video platforms in particular, recommendation systems influence not only which content is more likely to enter users' field of vision, but also the sequence, pace, and combination through which users encounter political information \citep{RN2,RN65}.

At the same time, the organizing role of recommendation systems does not necessarily result in equal content distribution across users \citep{RN60,RN45}. Prior research has shown that platform algorithms may reproduce and amplify existing social disparities in processes of ranking, filtering \citep{RN11,RN9}, and personalized distribution \citep{RN61}, with gender bias being particularly salient \citep{RN58,RN22,RN8}. For example, recommendation algorithms on Douyin have been found to preferentially promote anti-feminist and gender-stereotypical content \citep{RN21}. As recommendation systems increasingly mediate exposure to political information, gender should no longer be treated merely as a general form of bias on digital platforms, but rather as a critical entry point for understanding the consequences of algorithmic distribution. This raises an important question: do platform recommendation systems generate gendered asymmetries in the distribution of political information?

However, existing research on gender bias in recommendation systems has largely focused on output-level differences, such as disparities in exposure share, content diversity, or surface-level representational imbalance \citep{RN6,RN40,RN41}, while paying less attention to how content differences may be linked to structural differences in the information environments in which users are situated \citep{RN42}. In addition, although prior studies suggest that different groups may inhabit distinct information environments \citep{RN62}, it remains unclear whether such differences are primarily driven by users' pre-existing preferences and selective exposure, or by the active organizing role of platform recommendation mechanisms \citep{RN57}. To address this gap, this study focuses on the organizing role of recommendation systems. Specifically, by rigorously controlling for users' prior political preferences and behavioral trajectories, we examine whether platforms rely on gendered behavioral cues to direct users into structurally asymmetric information environments. Taking YouTube as a case, we conduct a controlled social-bot experiment to track homepage recommendation trajectories and analyze them from two complementary dimensions: allocative bias and structural bias. In doing so, we argue that gender bias in recommendation systems manifests not only as surface-level allocative bias, but also as deep-seated structural bias. Specifically, time-series analysis reveals that subsequent exposure pathways are continuously reinforced by existing structures. We further propose a simple collaborative-filtering model to explain one possible mechanism underlying the emergence of such gendered differences.

\section{Literature review}

\subsection{Algorithmic biases, and gender bias}

Recommendation algorithms have become ubiquitous in contemporary society and play a pervasive role in the automation of a wide range of tasks \citep{RN13}. However, systematic algorithmic bias may emerge at any stage of the algorithmic pipeline as it permeates everyday life, including data processing, algorithm development, and human--computer interaction \citep{RN79,RN80}. This means that algorithmic systems may produce asymmetric outcomes for different social groups through processes of classification, ranking, prediction, and content allocation. Prior research has shown that such bias can arise in a variety of contexts, including recruitment \citep{RN67}, advertising delivery \citep{RN34}, image search \citep{RN22}, and content recommendation \citep{RN19}. Among these forms of algorithmic bias, gender bias deserves particular attention. Recommendation systems may not only reproduce existing gender stereotypes, but also further reinforce gender bias through the differential allocation of information, opportunities, and visibility \citep{RN8}. For platform recommendation systems, whose core mechanisms are ranking and distribution, gender bias is most likely to be manifested through unequal visibility and differential content allocation.

In the field of news and political communication, gender bias in recommendation systems may be reflected not only in overall visibility, but also in issue association, narrative framing, and modes of presentation \citep{RN63}. Existing research shows that women politicians tend to receive less news attention overall \citep{RN59}, and they are also less likely to appear in core political issue areas such as the economy and national security \citep{RN63}. Even when they do enter news coverage, female candidates are more likely to be framed in terms of appearance, family, and personal traits, and such portrayals often undermine evaluations of their viability and public support. These disparities also extend to the visual level, as news images may construct politicians differently through facial expressions, facial features, and compositional choices \citep{RN64}.

As one of the most influential digital platforms, YouTube has become a major site through which the public accesses news and political information \citep{RN44}. This makes YouTube an especially important setting for examining gender bias in recommendation systems. As a content platform that relies heavily on sequential recommendation, political information exposure on YouTube is shaped less by one-time searches or incidental encounters than by continuous viewing and dynamic distribution. The recommendation system not only determines which videos are more likely to enter users' view, but also continually reshapes subsequent pathways of exposure, giving the entry, accumulation, and association of political content a distinctly processual character \citep{RN65}. Accordingly, gendered differences in distribution on YouTube should not be understood merely as asymmetries in the proportion of recommended content, but as a question of how the recommendation system configures the conditions of political exposure and further organizes political information environments. Although prior research has paid considerable attention to the role of YouTube's recommendation algorithm in filter bubbles \citep{WOS:001439101300001}, and ideological bias \citep{WOS:001128094700001}, relatively little is known about whether gendered behavioral cues may also direct different user profiles into differentiated political information environments.

\subsection{Allocative Bias}

Existing research on algorithmic fairness typically distinguishes between two core forms of algorithmic bias: representational bias and allocative bias \citep{RN70}. The former refers to the ways in which algorithms reinforce existing social stereotypes, demean particular groups, or render them invisible at the level of cultural representation—for example, by disproportionately associating women with subordinate occupations in image search results \citep{RN11}. The latter refers to unequal outcomes produced when automated systems allocate resources, opportunities, and information across different social groups \citep{RN69}. Although representational bias has attracted substantial attention in studies of vision and natural language models \citep{RN71,RN64}, the more central mechanism in recommendation systems, which function as infrastructures for content distribution, lies in the asymmetric allocation of information and opportunities for exposure. For this reason, the present study focuses on allocative bias.

On digital platforms, allocative bias is manifested most directly in asymmetries of content exposure and visibility \citep{RN72}. For instance, online advertising systems have been shown to deliver STEM-related content more frequently to male users \citep{RN34}. When this allocative logic extends into political information environments, it takes the form of recommendation systems distributing unequal political information to different groups. As prior research has shown, women politicians tend to receive less overall media attention than men in news coverage \citep{RN59}. Allocative bias is also reflected in the differentiated allocation of political content: even when different user profiles are exposed to similar overall quantities of political information, they may still be persistently directed toward different issue domains. For example, a meta-analysis by L. H. Sun found that women are significantly less likely than men to encounter core political content related to national security and the economy \citep{RN64}. At a deeper level, this allocative bias is rooted in the collaborative filtering logic of recommendation systems, through which algorithms capture small behavioral differences between users and progressively amplify them through ongoing feedback loops.

To systematically analyze this gendered form of allocative bias, it is necessary to specify its core dimensions of measurement.  Drawing on media diversity research and normative discussions of the democratic role of algorithms, this study operationalizes allocative bias along three dimensions: issue, ideology, and entity. Specifically, the distribution of issues is directly related to the frequency with which audiences encounter core political matters such as macroeconomic affairs, and thus to political empowerment \citep{RN75,RN74}. Cross-cutting exposure to ideology shapes whether audiences become trapped in homogeneous cognitive polarization, a risk that recommendation filtering on digital platforms may further intensify \citep{RN73}. The visibility of politicians of different genders, as well as related entities, in recommendation streams also has important implications for the pluralistic representation of political discourse \citep{RN59}. A comprehensive assessment of gendered allocative bias in recommendation systems therefore requires these three dimensions to be considered in an integrated framework.

Building on this distinction, the present study further refines allocative bias in content distribution into two closely related dimensions. The first concerns differences in the proportion and visibility of political content, that is, whether the recommendation system produces systematic differences in the overall share and diversity of political information encountered by different gender-coded behavioral profiles. The second concerns differences in the distribution of political content, that is, whether the system directs different profiles toward distinct content domains in terms of issue distribution, ideological orientation, and political entities. Through this refinement, gender bias in recommendation systems is no longer understood merely as a question of who sees more, but also of who is allocated what kinds of political content. Against this background, we propose the following research questions to examine whether recommendation outcomes differ across male-coded and female-coded profiles:

\textit{RQ1a}: Do different gender-coded behavioral profiles differ in the proportion and visibility of political content they receive?

\textit{RQ1b}: Do different gender-coded behavioral profiles differ in the distribution of issues, ideological orientations, and entities in the political content they receive?

\subsection{Structural Bias}

Beyond allocative bias at the level of content distribution, gender bias in recommendation systems may also be manifested in the systematic organization of information environments, namely, structural bias. Structural bias refers to persistent and systematic distributive skew that is generated endogenously by the metrics and rules embedded in ranking systems. Such skew does not result from isolated errors; rather, it arises from the algorithmic structure itself and its long-term prioritization of particular signals \citep{RN76}. Rooted in the strong optimization of user engagement, this form of bias often takes the form of algorithmic segregation in network topology \citep{RN62}. Under collaborative filtering, algorithms tend to produce clustered recommendation spaces characterized by dense internal connections and high within-group overlap through repeated similarity matching \citep{RN78,RN77}. As a result, even when different profiles are exposed to comparable amounts of political content, they may still occupy markedly different structural positions in the recommendation process, especially in terms of connection density and community boundaries. In this way, recommendation systems can reshape users' pathways of information exposure at a deeper structural level. Such structural differentiation may lock certain groups into highly homogeneous information clusters, thereby limiting exposure to heterogeneous information and reinforcing existing cognitive biases.

However, existing research on algorithmic segregation has focused primarily on ideology-driven differentiation, that is, on how algorithms contribute to political polarization or echo chambers \citep{RN66,RN67}. By contrast, there has been little dynamically grounded empirical research on whether recommendation systems may systematically organize and segregate audiences at the level of underlying network topology on the basis of gendered behavioral cues. Because structural bias is both latent and self-reinforcing, content analysis alone is insufficient to capture its dynamic evolution. To address this limitation, the present study introduces SNA (Social Network Analysis) and quantifies differences in topological indicators such as density, modularity, and community overlap across recommendation networks associated with different gender-coded profiles. In doing so, it aims to examine whether recommendation systems generate structural bias toward different users at the level of their underlying organizational logic. On this basis, we propose the following research question:

\textit{RQ2}: Does YouTube's recommendation system organize structurally asymmetric political information environments around different gender-coded behavioral profiles?

Building on the identification of such structural bias, the present study further examines how it operates as a dynamic process. Specifically, the recommendation process is conceptualized as a feedback chain of exposure, click, and re-exposure, and we further test whether existing community boundaries and structural positions continue to exert an organizing influence on subsequent clicks and renewed exposure. Through this analysis, the study seeks to determine whether structural bias in recommendation systems is merely a static form of network differentiation, or whether it is continuously reinforced through feedback loops and, in turn, shapes users' subsequent pathways of information exposure.

\section{Materials and Methods}

By constructing controlled social bots, this study investigates whether the YouTube recommendation system generates differentiated political information environments for distinct gendered behavioral profiles. A total of 160 virtual accounts were constructed for the experiment, consisting of 80 male-coded and 80 female-coded behavioral profiles. Here, "male" and "female" refer not to users' actual demographic identities but to gendered behavioral signals constructed based on prevalent category preferences on the platform. Drawing upon the identification of gender distributions across official YouTube categories by Mike Thelwall and David Foster \citep{RN4,richmond2016openslate}, and incorporating YouTube audience demographics published by OpenSlate, the male-coded behavioral categories were defined as \textit{Autos \& Vehicles}, \textit{Sports}, and \textit{Gaming}, whereas the female-coded behavioral categories were designated as \textit{Howto \& Style} and \textit{People \& Blogs}. The dynamic preference category was uniformly set to \textit{News \& Politics} across all accounts. Apart from these behavioral tags, all accounts were maintained under identical conditions in terms of creation time, operating environment, viewing protocols, and interaction steps.

The experiment consisted of two stages: the training phase and the testing phase. During the training phase, each account viewed 10 pre-selected "seed videos" at an $8:2$ ratio to establish a baseline profile. Specifically, $80\%$ of these videos were drawn from gendered behavioral categories, while $20\%$ belonged to the \textit{News \& Politics} category. The automated program required a video completion rate of over $80\%$, with a maximum viewing duration capped at $2{,}700$ seconds per video. This procedure activated the platform's profiling mechanism while limiting potential distortion from abnormal dwell times. In the testing phase, each account performed $150$ consecutive interaction steps along the platform's recommendation stream. At each step, the program recorded the recommended videos and executed clicks and views according to a pre-defined rule: $80\%$ of clicks targeted gendered behavioral categories, and $20\%$ targeted \textit{News \& Politics}. The time intervals between adjacent clicks were configured to follow a power-law distribution \((\alpha=-1)\), approximating the bursty and non-linear browsing rhythms characteristic of real-world users on digital platforms.

\subsection{Methods Study 1}

\subsubsection{Allocative Bias: Content Coding and Measurement}

To characterize the recommendation content received by different accounts, we adopted a two-level analytical strategy. First, using the full 150-step recommendation trajectories, we examined the dynamic evolution of political exposure across profile groups. Second, because recommendation patterns are more likely to stabilize in later stages of repeated interactions, we used the final 50 steps as the main analytical window to compare the proportion of political content and content diversity across groups.Overall content diversity was measured using Shannon entropy based on YouTube video categories:
\begin{equation}
H(X) = - \sum_{i=1}^{n} p_i \log_2 p_i,
\end{equation}
where \(n\) is the number of categories observed in the exposure set, and \(p_i\) is the proportion of videos in category \(i\) for a given account within the analytical window. We further calculated structural diversity by regrouping all videos into three broader classes: target interest (\textit{Interest}), political content (\textit{News \& Politics}), and background content (\textit{Others}). Structural diversity was computed using the same entropy formula applied to these three aggregated categories. As a robustness check, we also repeated the analyses using alternative terminal windows (last 30, last 40, last 50, last 60, and last 70 steps), with the results reported in the Appendix.

From the political videos identified in the final 50 recommendation steps, we compared differences across profile groups along three dimensions: issues, ideology, and entities. Political videos were annotated using an LLM-assisted hybrid coding framework that combined automated classification with manual validation. For issue analysis, we followed the Comparative Agendas Project (CAP) framework and assigned each video to one core issue category \citep{cap_master_codebook}. Each account was then represented as an issue-distribution vector, and similarity between accounts was measured using cosine similarity:
\begin{equation}
\mathrm{CosSim}(x,y)=\frac{\sum_{i=1}^{K} x_i y_i}{\sqrt{\sum_{i=1}^{K} x_i^2}\sqrt{\sum_{i=1}^{K} y_i^2}},
\end{equation}
where \(K\) denotes the number of core issue categories, and \(x_i\) and \(y_i\) denote the relative share of issue \(i\) for two accounts. We also used Shannon entropy to assess differences in issue diversity across groups. For ideology, videos were classified as \textit{Right Bias}, \textit{Left Bias}, or \textit{Least Biased}. For entity analysis, we extracted high-frequency person and organization entities from video titles to identify the political actors emphasized in different recommendation trajectories.

GPT-5 mini was used for the initial classification of issues, ideological orientations, and entities. We then randomly sampled \(10\%\) of the model outputs for manual coding and compared the human-coded results with the model predictions. Issue classification achieved an accuracy of \(89.4\%\) with a Cohen's kappa of 0.81, and ideology classification achieved an accuracy of \(93.1\%\) with a Cohen's kappa of 0.76. For entity recognition, precision, recall, and F1 were \(90.7\%\), \(87.2\%\), and \(88.9\%\), respectively.

\subsubsection{Structural Bias: Network Construction and Analysis}

To examine structural differences beyond allocative bias in content distribution, we constructed account co-exposure networks from exposure records within each analytical window. Nodes represent accounts, and edge weights represent the number of videos shared by a pair of accounts. Let \(X=(x_{ik})\) denote a binary account--video matrix, where \(x_{ik}=1\) if account \(i\) was exposed to video \(k\), and \(x_{ik}=0\) otherwise. The co-exposure weight between accounts \(i\) and \(j\) was defined as:
\begin{equation}
w_{ij}=\sum_k x_{ik}x_{jk}.
\end{equation}
An undirected edge was retained only when the co-exposure weight exceeded a specified threshold $\theta$ (i.e., $w_{ij} > \theta$). For our primary analysis, we set $\theta = 20$. To ensure the robustness of our network structure, alternative thresholds (e.g., $\theta = 10$, $\theta = 15$, $\theta = 25$, $\theta = 30$) were also evaluated.

Based on the co-exposure network, we computed network density, the weighted average clustering coefficient, and weighted modularity to characterize the organization of political information environments across profile groups. Network density was defined as:

\begin{equation}
D=\frac{2E}{N(N-1)},
\end{equation}

where \(E\) is the number of observed edges and \(N\) is the number of nodes. The weighted average clustering coefficient was computed as:

\begin{equation}
C_w=\frac{1}{N}\sum_{i=1}^{N} C_i^{w}
=\frac{1}{N}\sum_{i=1}^{N}\frac{1}{s_i(k_i-1)}
\sum_{j,h}\frac{w_{ij}+w_{ih}}{2}\,a_{ij}a_{ih}a_{jh},
\end{equation}

where \(k_i\) is the degree of node \(i\), \(s_i=\sum_j w_{ij}\) is its weighted degree, and \(a_{ij}\) is the adjacency matrix. To detect community structure, we applied the Louvain algorithm and calculated weighted modularity:

\begin{equation}
Q=\frac{1}{2m}\sum_{i,j}\left(w_{ij}-\gamma\frac{s_i s_j}{2m}\right)\delta(c_i,c_j),
\end{equation}

where \(m=\frac{1}{2}\sum_i s_i\), \(c_i\) denotes the community assignment of node \(i\), and \(\delta(c_i,c_j)\) equals 1 when two nodes belong to the same community and 0 otherwise. Higher modularity indicates stronger partitioning of the recommendation space into relatively independent modules. The main structural analysis focused on political exposures in the final 50 recommendation steps. We also constructed co-exposure networks for the first 50 and last 50 steps and compared modularity across windows to assess whether structural differentiation increased over time. In addition, we compared the dominant issue and ideological composition of detected communities to describe the informational characteristics associated with different community boundaries.

After identifying structural differences across gender-coded behavioral profiles, we examined whether these structural boundaries continued to shape subsequent information exposure through the feedback chain of exposure, click, and re-exposure. The recommendation process was divided into consecutive stages, and political exposure and click records in each stage were mapped into issue-distribution vectors. For each account, we defined four reference structures: self, within-community, within-group but outside-community, and out-group. Except for the self-level reference, all reference structures were constructed as average vectors of the corresponding sets of other accounts, excluding the focal account itself.

In the descriptive analysis, we compared the recommendation--click--re-recommendation process at both the video-ID level and the issue level. At the ID level, we calculated Jaccard overlap between accounts, as well as between each account and the corresponding reference structures at different levels, based on sets of video IDs:
\begin{equation}
J(A,B)=\frac{|A\cap B|}{|A\cup B|},
\end{equation}
where \(|A\cap B|\) denotes the number of video IDs shared by the two sets, and \(|A\cup B|\) denotes the number of distinct video IDs in their union. This measure captures overlap in specific content between recommendation and click records. At the issue level, we calculated cosine similarity between exposure issue vectors and click issue vectors to assess which structural level subsequent clicks or exposures were closest to in terms of issue composition. Specifically, we compared an account's click outcomes at stage \(t+1\) with the exposure structures at different levels in stage \(t\), and an account's exposure outcomes at stage \(t+1\) with the click structures at different levels in stage \(t\). This allowed us to assess whether the recommendation--click--re-recommendation process primarily unfolded along community boundaries, and whether within-community structure was more closely associated with subsequent behavior and exposure outcomes than other structural levels.

Building on the descriptive comparisons, we estimated two sets of lagged regression models to assess the relative explanatory power of structural boundaries in the feedback process. The first set examined how exposure structures in the previous stage shaped the issue composition of clicks in the subsequent stage. Let \(s_{i,t+1,\ell}\) denote the relative share of issue \(\ell\) in the click issue vector of account \(i\) at stage \(t+1\). The model is specified as:
\begin{equation}
s_{i,t+1,\ell}
=
\beta_1 e^{\mathrm{self}}_{i,t,\ell}
+
\beta_2 e^{\mathrm{comm}}_{i,t,\ell}
+
\beta_3 e^{\mathrm{in/out}}_{i,t,\ell}
+
\beta_4 e^{\mathrm{outgroup}}_{i,t,\ell}
+
\alpha_{\ell}
+
\lambda_t
+
\eta_g
+
\varepsilon_{i,t,\ell},
\end{equation}
where \(e^{\mathrm{self}}_{i,t,\ell}\), \(e^{\mathrm{comm}}_{i,t,\ell}\), \(e^{\mathrm{in/out}}_{i,t,\ell}\), and \(e^{\mathrm{outgroup}}_{i,t,\ell}\) denote the relative share of issue \(\ell\) in the stage-\(t\) exposure structure at four levels: the focal account itself, the focal account's community, accounts within the same profile group but outside the focal community, and accounts in the out-group, respectively. \(\alpha_{\ell}\), \(\lambda_t\), and \(\eta_g\) denote issue, stage, and group fixed effects.

The second set examined the feedback effect of click issue structures in the previous stage on exposure structures in the subsequent stage. Let \(e_{i,t+1,\ell}\) denote the relative share of issue \(\ell\) in the exposure issue vector of account \(i\) at stage \(t+1\). The model is specified as:
\begin{equation}
e_{i,t+1,\ell}
=
\beta_1 c^{\mathrm{self}}_{i,t,\ell}
+
\beta_2 c^{\mathrm{comm}}_{i,t,\ell}
+
\beta_3 c^{\mathrm{in/out}}_{i,t,\ell}
+
\beta_4 c^{\mathrm{outgroup}}_{i,t,\ell}
+
\alpha_{\ell}
+
\lambda_t
+
\eta_g
+
\varepsilon_{i,t,\ell},
\end{equation}
where \(c^{\mathrm{self}}_{i,t,\ell}\), \(c^{\mathrm{comm}}_{i,t,\ell}\), \(c^{\mathrm{in/out}}_{i,t,\ell}\), and \(c^{\mathrm{outgroup}}_{i,t,\ell}\) denote the relative share of issue \(\ell\) in the stage-\(t\) click structure at the same four structural levels.

By combining ID-level overlap analysis, issue-level similarity comparisons, and lagged regression models, we assessed whether structural bias in the recommendation system was merely a static form of network differentiation or instead evolved into a relatively stable path dependence through repeated interaction.

\begin{figure}[pos = htbp]
    \centering
    \includegraphics[width= 0.8\textwidth]{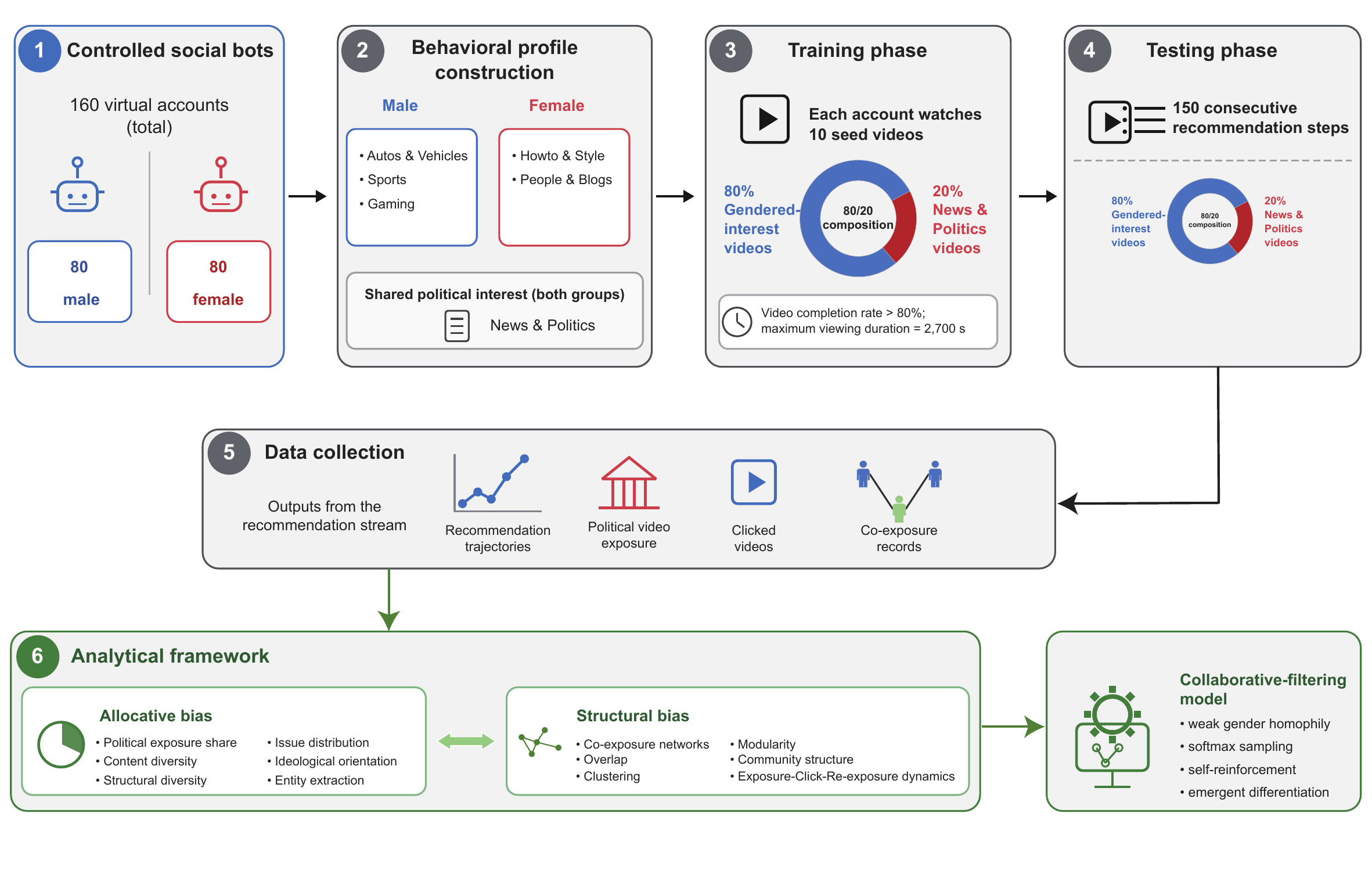}
    \caption{Methodological workflow of the study. The figure summarizes the construction of controlled social bots, the operationalization of gender-coded behavioral profiles, the training and testing phases, the collection of recommendation trajectory data, the analytical framework for allocative and structural bias, and the collaborative-filtering simulation used in Study 2.}
    \label{fig:fig1}
\end{figure}

\subsection{Methods Study 2}

To provide a supplementary explanation for how differentiated political information environments may emerge in recommendation systems, we developed a simple model from the perspective of collaborative filtering. Study 2 was not intended to reproduce the actual implementation of YouTube's recommendation algorithm or to directly identify the platform's underlying mechanism. Instead, it used numerical simulation to examine whether a recommendation system with a weak gender-homophily preference imposed on gender-coded behavioral profiles could generate both content-level and structural differentiation over repeated interactions, even in the absence of pre-specified initial political preferences.

The model consists of a population of agents with gender-coded profile labels embedded in a multidimensional political issue space. At time step \(t\), the cognitive state of agent \(i\) is defined as a 21-dimensional exposure-salience vector \(\mathbf{x}_i(t)\), representing the relative weights of issues pushed by the system and effectively received by the agent. To exclude the prior assumption that polarization arises from initial differences in political preferences, all initial state vectors \(\mathbf{x}_i(0)\) were randomly drawn from a uniform distribution \(U(0,1)\) at \(t=0\) and then normalized such that
\begin{equation}
\sum_{k=1}^{21} x_{i,k}(0)=1.
\end{equation}
This initialization creates a starting condition without group-level political differences across gender-coded labels. The system also records each agent's gender label \(g_i \in \{m,f\}\).

The model does not impose any predefined mapping between gender and specific political issues. Instead, content distribution is driven purely by similarity-based collaborative filtering. At each time step \(t\), the system computes a composite similarity score \(S_{ij}\) between agent \(i\) and another agent \(j\):
\begin{equation}
S_{ij}=\cos\!\big(\mathbf{x}_i(t),\mathbf{x}_j(t)\big)+\beta \cdot \delta(g_i,g_j),
\end{equation}
where the first term denotes the cosine similarity between two state vectors in the 21-dimensional issue space; \(\beta\) captures the strength of the recommendation system's gender-homophily preference, that is, the additional weight assigned to agents with the same gender-coded profile label when the system evaluates potential similarity; and \(\delta(g_i,g_j)\) is the Kronecker function, which equals 1 when the two agents share the same gender label and 0 otherwise.

Recommendation probabilities were generated using a nonlinear Softmax-based sampling rule rather than a simple linear proportional rule. The probability that content generated from the state of agent \(j\) is recommended to agent \(i\), denoted by \(P_{ij}\), is defined as
\begin{equation}
P_{ij}=\frac{\exp(S_{ij}/\tau)}{\sum_{k \ne i}\exp(S_{ik}/\tau)},
\end{equation}
where \(\tau\) is a temperature parameter. Under highly homogeneous initial conditions, cosine-distance differences between agents are extremely small. A relatively low value of \(\tau\) can substantially amplify the structural differences induced by the weak bias term \(\beta\), making the algorithm much more likely to recommend content sources associated with agents sharing the same label, thereby creating an early tendency toward segregation in the collaborative-filtering process.

After receiving the set of recommended content items \(\Omega_i\) sampled by the system according to \(P_{ij}\), agent \(i\) updates its salience vector. Conventional bounded-confidence models generally assume convergence toward the mean of the surrounding recommendation environment. In a multidimensional space, however, such a smoothing mechanism can easily drive the system toward central collapse. To avoid this, the model builds on the Hegselmann--Krause (HK) framework but replaces linear averaging with a nonlinear self-reinforcement rule based on the Hadamard product. The updating equation for agent \(i\) is
\begin{equation}
\mathbf{x}_i(t+1)=\mathrm{Norm}\left(\mathbf{x}_i(t)+\alpha \sum_{j \ne i} P_{ij}\,\mathbf{x}_j(t)\odot \mathbf{x}_i(t)\right),
\end{equation}
where \(\mathbf{x}_j(t)\) is the current state vector of the recommended source agent \(j\), \(\odot\) denotes element-wise multiplication, \(\alpha\) is the learning rate, and \(\mathrm{Norm}(\cdot)\) denotes L1 normalization applied to the updated vector to preserve its probabilistic interpretation.

\FloatBarrier

\section{Results}

A total of 160 controlled social bots were deployed in this study, including 80 male-coded profiles and 80 female-coded profiles. From the continuous recommendation trajectories of these accounts, we collected 509,336 recommendation exposures, among which 78,728 were identified as political content exposures. The results are reported in three parts. We first examine gender bias in the recommendation system at the content level, then at the structural level, and finally use a simplified model to discuss a possible generative mechanism from the perspective of collaborative filtering.

\subsection{Allocative Bias in Political Content Exposure}

\begin{figure}[pos = htbp]
    \centering
    \begin{minipage}[b]{0.47\textwidth}
        \centering
        \includegraphics[width=\textwidth]{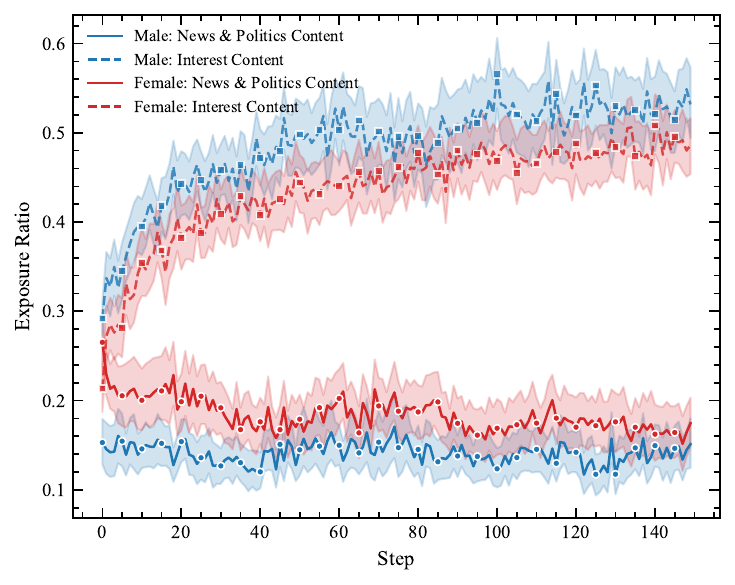}
        \vspace{0.5em}
        \centerline{(a)}
    \end{minipage}
    \hfill
    \begin{minipage}[b]{0.48\textwidth}
        \centering
        \includegraphics[width=\textwidth]{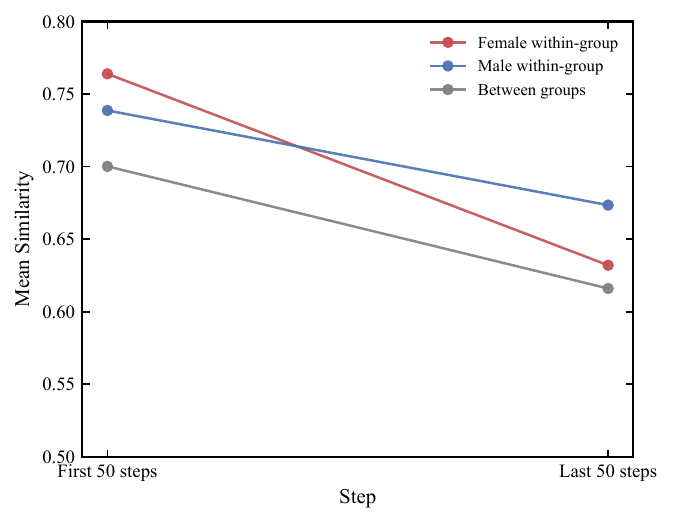}
        \vspace{0.5em}
        \centerline{(b)}
    \end{minipage}
    \caption{(a) Dynamic trajectories of content exposure across gender-coded profiles over 150 consecutive recommendation steps. Solid lines indicate the proportion of political news exposure, and dashed lines indicate the proportion of target-interest exposure in the recommendation streams of the two profile groups. Female-coded and male-coded profiles are shown in red and blue, respectively. Shaded areas denote 95\% confidence intervals. (b) Evolution of pairwise content similarity across different stages of the experiment, showing the mean pairwise similarity within groups (female group, red; male group, blue) and between groups (gray).}
    \label{fig:figure2}
\end{figure}

The results indicate differences in content allocation across gender-coded profiles in YouTube's recommendation system. Among the \(78{,}728\) political content exposures identified from the full recommendation trajectories, \(34{,}176\) were received by male-coded accounts (\(43.41\%\)) and \(44{,}552\) by female-coded accounts (\(56.59\%\)). As shown in Fig.~\ref{fig:figure2}(a), female-coded profiles consistently exhibited a higher proportion of political content exposure than male-coded profiles in the absence of gender-differentiated prior political interaction. For a more stable comparison, we focused on the final 50 recommendation steps. Within this window, the mean proportion of political news videos was \(17.2\% \pm 10.1\%\) for female-coded accounts, compared with \(13.4\% \pm 7.8\%\) for male-coded accounts, and the difference was statistically significant (\(p < 0.01\)). Comparable results for the last 30--70 steps are reported in Appendix Table~\ref{tab:appendix1}. A different pattern emerged for exposure diversity. The male group showed significantly higher overall entropy than the female group (\(2.79 \pm 0.37\) vs.\ \(2.68 \pm 0.27\), \(p < 0.05\)), indicating broader coverage of fine-grained categories. By contrast, structural entropy was significantly lower in the male group than in the female group (\(1.32 \pm 0.21\) vs.\ \(1.40 \pm 0.15\), \(p < 0.01\)); corresponding results for the last 30--70 steps are reported in Appendix Table~\ref{tab:appendix2}. Overall, these results suggest a measurable divergence in late-stage content allocation: female-coded accounts were more likely to encounter political content, whereas male-coded accounts were more likely to remain concentrated in their preexisting interest categories. However, viewed against the platform's overall recommendation stream, the magnitude of this difference was relatively modest.

However, similar overall proportions of political exposure do not imply that different profiles were allocated the same kinds of political content. We first compared the similarity of political issue distributions across the earlier and later analytical windows. As shown in Fig.~\ref{fig:figure2}(b), within-group similarity was consistently higher than between-group similarity, indicating that the recommendation system allocated different political content to different gender-coded profiles. Notably, the more pronounced decline in within-group similarity for the female-coded group in the later stage indicates that content similarity within the female-coded group was less concentrated than within the male-coded group. In addition, political issue diversity in the final 50 steps was significantly higher for female-coded accounts than for male-coded accounts (\(2.11 \pm 0.34\) vs.\ \(1.91 \pm 0.54\); \(p < 0.05\)). This pattern likewise suggests that the platform was more likely to concentrate male-coded users' political exposure within a narrower set of issue domains, while placing female-coded users in a broader and more diffuse issue environment.

\begin{figure}[pos = htbp]
    \centering
    \includegraphics[width=0.7\textwidth]{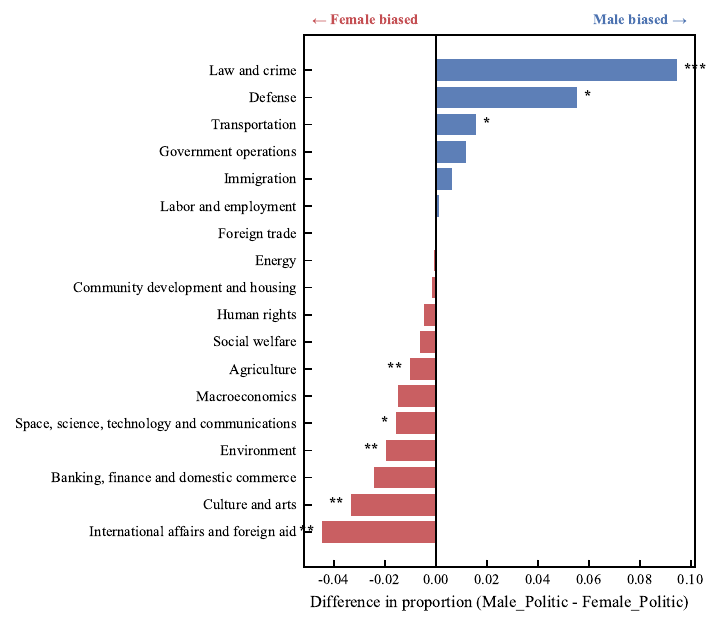}
    \caption{Differences in the distribution of specific political issues across gender-coded profiles. The horizontal bar chart displays differences in the recommendation share of specific political issues between female-coded and male-coded accounts. Positive values (blue bars) indicate higher exposure shares among male-coded profiles, whereas negative values (red bars) indicate higher exposure shares among female-coded profiles. Asterisks indicate statistical significance (*\(p<0.05\), **\(p<0.01\), ***\(p<0.001\)).}
    \label{fig:figure3}
\end{figure}

Having established differences in issue diversity, we further compared the relative proportions of specific political issues in the exposure content in order to identify substantive differences in these gendered information environments. For each political issue, we conducted between-group comparisons using independent-samples t-tests to assess whether the mean exposure proportion differed significantly between male-coded and female-coded profiles. As shown in Fig.~\ref{fig:figure3}(a), the distribution of issue proportions reveals an asymmetric pattern in the system's allocation of fine-grained political issues. The political recommendation stream for male-coded profiles was highly concentrated in a small number of hard issues related to domestic order and infrastructure, with higher exposure shares than female-coded profiles in both \textit{Law and crime} (\(p < 0.001\)) and \textit{Defense} (\(p < 0.05\)). By contrast, female-coded accounts received significantly greater exposure to a broader range of issues with more macro-level and lifestyle-related orientations. In particular, female-coded profiles showed higher exposure shares than male-coded profiles in \textit{International affairs and foreign aid} (\(p < 0.01\)) and \textit{Culture and arts} (\(p <0.01\)); see Appendix Table~\ref{tab:appendix3} for details. Taken together, these findings suggest that the recommendation system did not push the two groups toward opposite ends of a polarized ideological spectrum. Rather, it implemented an asymmetric issue allocation strategy: male-coded users were steered toward a narrower set of confrontational domestic-order issues, whereas female-coded users were exposed to a broader informational environment populated by more multidimensional, moderate, and establishment-oriented macro and public-policy issues.

Beyond issue differentiation, the recommendation system also exhibited significant differences at the ideological level. The mean proportion of neutral content received by female-coded accounts reached \(70.9\% \pm 18.1\%\), significantly higher than that of male-coded accounts (\(63.3\% \pm 20.3\%\); \(p = 0.02\)). Notably, the distribution of explicitly ideological content revealed a pattern that ran counter to conventional expectations. There was no statistically significant difference between the two profile groups in exposure to right-leaning content (\textit{Right Bias}; female: \(12.5\% \pm 12.0\%\), male: \(11.5\% \pm 11.3\%\), \(p = 0.50\)). Instead, the more pronounced difference emerged in left-leaning content (\textit{Left Bias}): male-coded accounts received a mean proportion of \(22.9\% \pm 18.8\%\), higher than the \(18.0\% \pm 14.2\%\) received by female-coded accounts, with the difference marginally significant(\(p = 0.07\)). Overall, ideological allocation in the recommendation system did not take the form of a symmetrical split between the left and right poles. Rather, it resembled an asymmetric configuration in which female-coded profiles were stabilized within a more neutral and lower-conflict informational zone, whereas male-coded profiles received a higher proportion of polarized content exposure.

At the entity level, the results provide additional evidence that the recommendation system made different political actors and institutions salient across profile groups. The entity-level results do not suggest a simple gendered split between right-wing political actors and mainstream economic commentators. Instead, both profile groups prominently featured major political figures such as Trump and Putin, while differing in the relative salience of other entities. Male-coded profiles showed a relatively stronger presence of  state-power related entities, including \textit{ICE}, \textit{DOJ}, and \textit{MAGA}, alongside major news programs such as \textit{FRONTLINE}, \textit{BBC}, and \textit{60 Minutes}. Female-coded profiles, by contrast, showed more frequent references to China-related political actors and institutions, as well as several institutional media and public-affairs outlets, including \textit{Business Insider}, \textit{PBS NewsHour}, and \textit{Amanpour and Company}. These patterns suggest that platform recommendations varied not only in the topics they presented, but also in the political actors and institutions through which those issues were framed. See Appendix Fig.~\ref{fig:appendix1} for details.

\subsection{Structural Bias in Political Information Environments}

\begin{figure}[pos = htbp]
    \centering
    \includegraphics[width=0.75\textwidth,height=0.28\textheight,keepaspectratio]{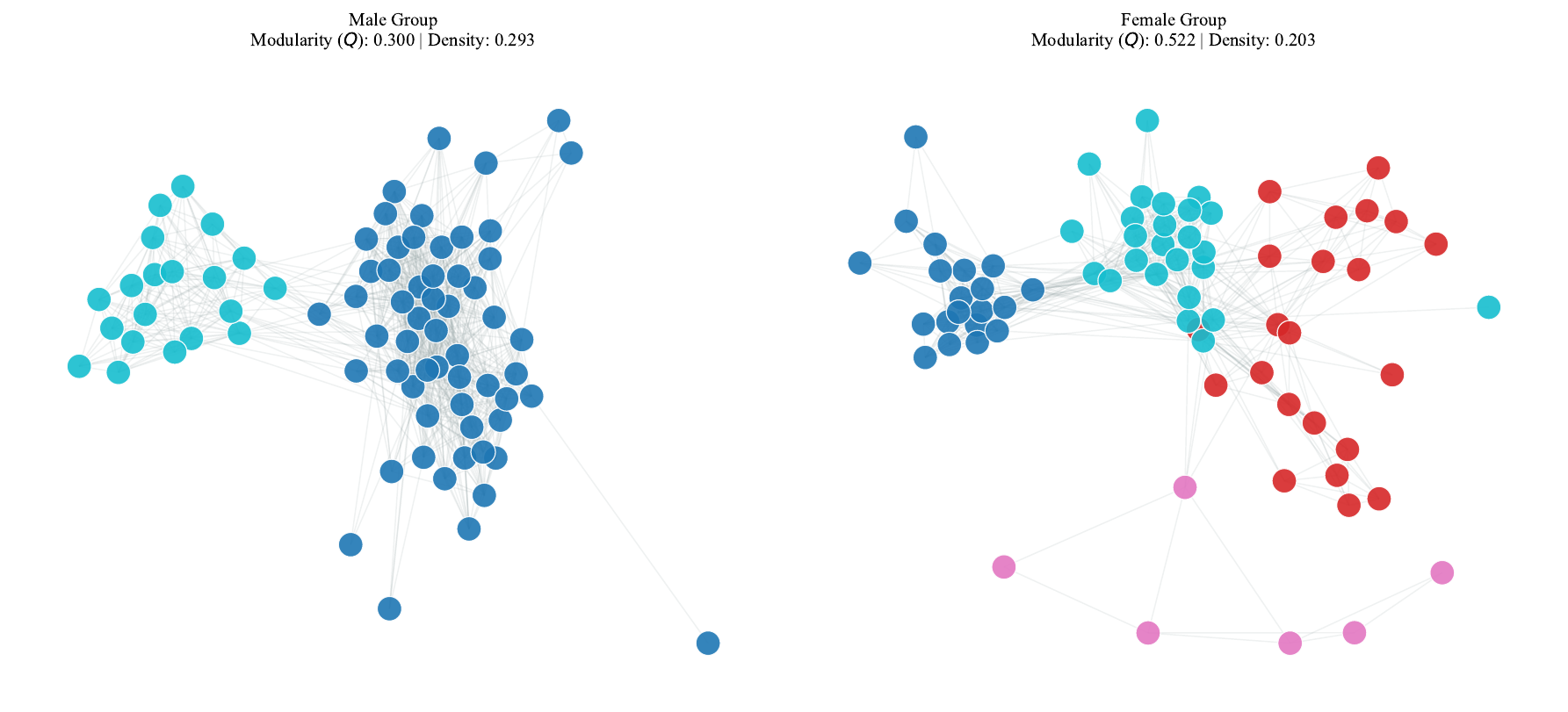}
    \centerline{\makebox[0.45\textwidth][c]{(a)} \hspace{0.05\textwidth} \makebox[0.45\textwidth][c]{(b)}}
    \vspace{0.6em}
    \includegraphics[width=0.75\textwidth,height=0.28\textheight,keepaspectratio]{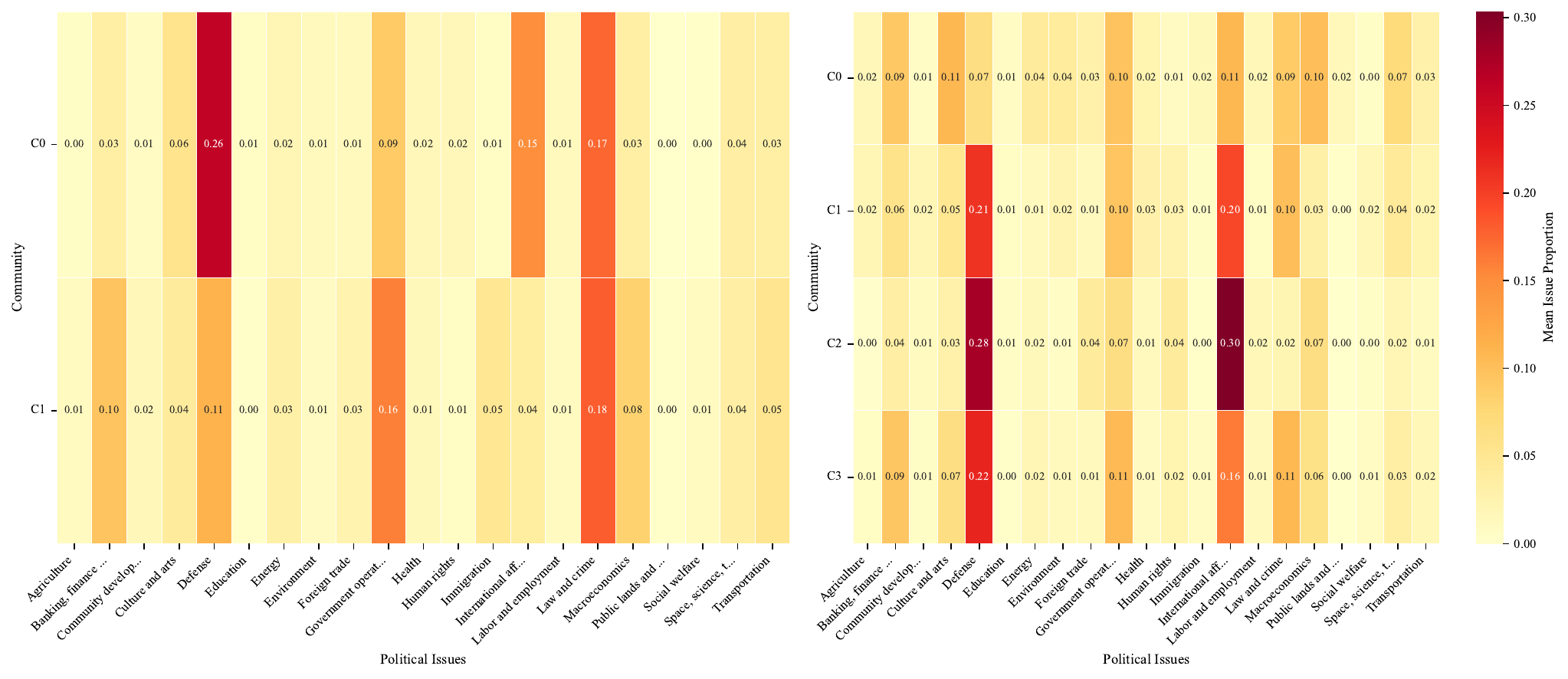}
    \centerline{\makebox[0.45\textwidth][c]{(c)} \hspace{0.05\textwidth} \makebox[0.45\textwidth][c]{(d)}}
    \vspace{0.6em}
    \includegraphics[width=0.75\textwidth,height=0.28\textheight,keepaspectratio]{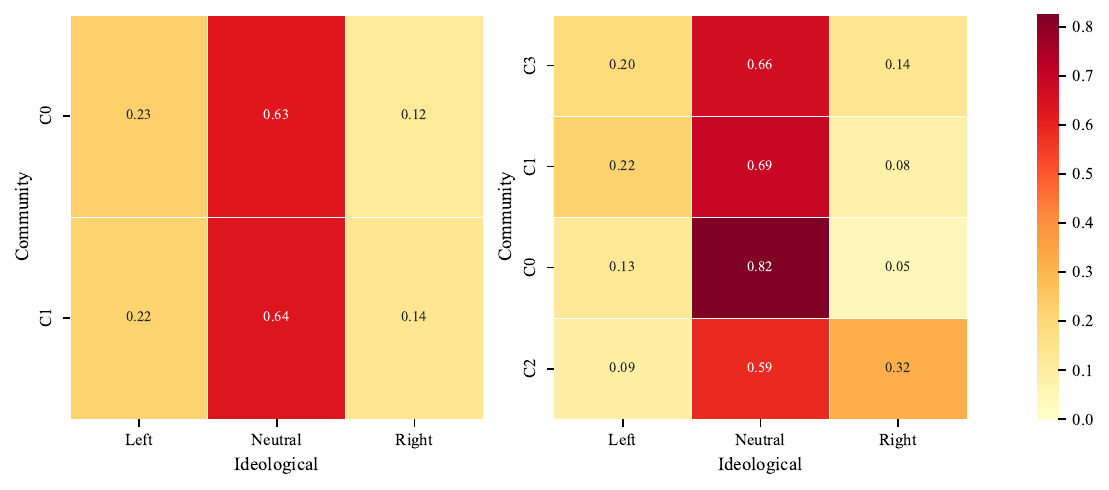}
    \centerline{\makebox[0.45\textwidth][c]{(e)} \hspace{0.05\textwidth} \makebox[0.45\textwidth][c]{(f)}}
    \vspace{0.6em}
 
    \caption{Community structure and within-community content composition across gender-coded profiles. Panels (a) and (b) present the community partition results of account co-exposure networks for male-coded accounts and female-coded accounts, respectively. Each node represents an account, node colors indicate communities identified by the community detection algorithm, and edges denote co-exposure links between accounts based on shared political video exposure. Panels (c) and (d) show the mean issue distribution within each detected community for male-coded and female-coded accounts, respectively. Panels (e) and (f) show the ideological distribution within each detected community for male-coded and female-coded accounts, respectively. Darker colors indicate a higher relative share of the corresponding issue or ideological category within a given community.}

    \label{fig:figure4}
    
\end{figure}

From the account co-exposure networks, gender bias in the recommendation system extended beyond content allocation and were reflected in the organization of political information environments. Both groups showed identifiable community structure, but the female-coded network exhibited clearer community differentiation and greater internal heterogeneity, whereas the male-coded network was more concentrated and issue-homogeneous. Subgroups in the male-coded network clustered mainly around domestic-order and security-related issues, especially \textit{Law and crime} and \textit{Defense}, while communities in the female-coded network formed relatively distinct clusters around \textit{Culture and arts}, \textit{International affairs and foreign aid}, and \textit{Macroeconomics} (see Fig.~\ref{fig:figure4}). Consistent with this pattern, modularity was higher in the female-coded group than in the male-coded group in both the first 50 and the last 50 recommendation steps, and it increased more markedly over the course of recommendation (see Appendix Fig.~\ref{fig:appendix2} for details).

The account co-exposure networks constructed from political exposures in the final 50 recommendation steps further revealed clear topological differences between the two information environments. Although the two groups had a similar number of nodes (female: 75; male: 78), the male-coded network exhibited a greater number of co-exposure edges, higher network density (\(0.29\) vs.\ \(0.20\)), and a higher average clustering coefficient (\(0.21\) vs.\ \(0.19\)) than the female-coded network. This indicates that male-coded accounts were more likely to encounter the same political content repeatedly and to form local structures characterized by high overlap and strong cohesion. By contrast, although female-coded accounts were exposed to a broader range of content, the degree of shared exposure within the group was weaker and the overall network was more diffuse. Further permutation tests based on 1,000 label reshufflings showed that the higher network density observed in the male-coded group was statistically significant (\(p = 0.003\)), whereas the difference in the average clustering coefficient was not (\(p = 0.635\)). Modularity was higher in the female-coded group than in the male-coded group, but this difference reached only marginal significance (\(p = 0.076\)). In other words, the recommendation system did not direct the two profile groups into a homogeneous political information space, but instead organized two distinct structural forms: a more concentrated and shared network for male-coded profiles, and a more diffuse and differentiated one for female-coded profiles (see Appendix Figure ~\ref{fig:appendix2_all} for results at $\theta = 10$, $\theta = 15$, $\theta = 25$, and $\theta = 30$).

\FloatBarrier

%
%
%
%
%
%

After identifying this structural asymmetry, we further examined whether it continued to shape subsequent exposure pathways through the feedback chain of exposure, click, and re-exposure. The results showed strong continuity in late-stage communities, indicating that these communities were not incidental graph structures but stable boundaries that remained closely linked to earlier network configurations (mean community continuity = \(0.995\) for the male-coded group and \(0.8825\) for the female-coded group). Further hierarchical comparisons showed that, in the exposure \(\rightarrow\) subsequent click stage, an account's own prior exposure path was the strongest predictor of subsequent clicks, although within-community structure still provided significant additional explanatory power (see Appendix Table~\ref{tab:appendix4} and Appendix Table~\ref{tab:appendix5}). By contrast, in the click \(\rightarrow\) subsequent exposure stage, within-community click structure emerged as the strongest predictive level, substantially outperforming self, within-group but outside-community, and out-group structures (see Appendix Table~\ref{tab:appendix4} and Appendix Table~\ref{tab:appendix6}). A similar pattern was observed at the video-ID level (see Appendix Fig.~\ref{fig:appendix3}).

\begin{figure}[pos = htbp]
    \centering
    \includegraphics[width=\textwidth]{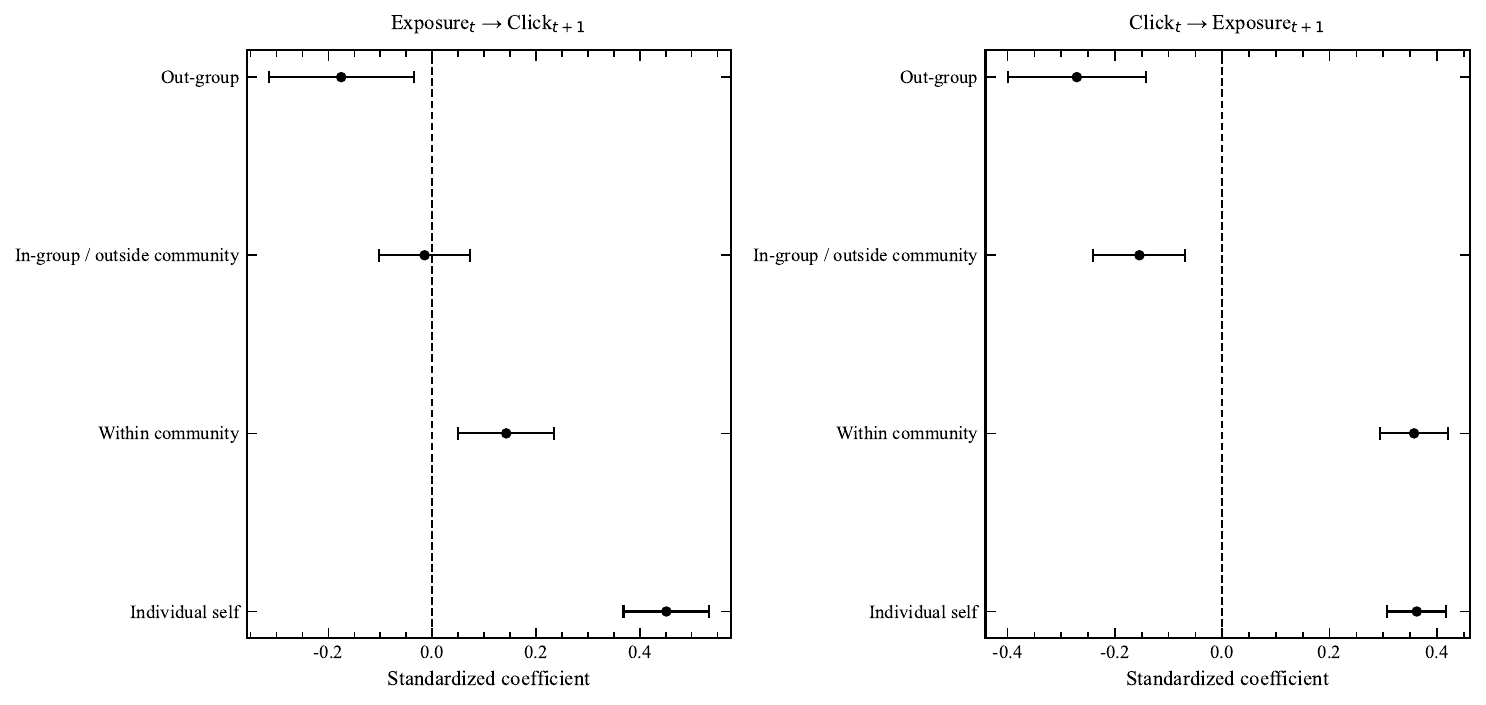}
    \centerline{\makebox[0.45\textwidth][c]{(a)} \hspace{0.05\textwidth} \makebox[0.45\textwidth][c]{(b)}}    
    \caption{Standardized coefficients from lagged regression models. The left panel shows the effects of exposure structures at stage \(t\) on click issue composition at stage \(t+1\), and the right panel shows the effects of click structures at stage \(t\) on exposure issue composition at stage \(t+1\). Points represent standardized coefficients and horizontal bars denote 95\% confidence intervals. Positive coefficients indicate greater similarity to the corresponding structural level in the subsequent stage.}

    \label{fig:figure5}
\end{figure}

The lagged regression results further support this interpretation. Fig.~\ref{fig:figure5} shows that, in the exposure \(\rightarrow\) subsequent click model, prior self-exposure exerted the strongest effect (\(\beta = 0.45\), \(p < 0.001\)), while within-community exposure structure also showed a significant positive effect (\(\beta = 0.14\), \(p < 0.001\)). In the click \(\rightarrow\) subsequent exposure model, within-community click structure played a role comparable to prior self-clicks, with both positively predicting subsequent exposure (\(\beta = 0.36\), \(p < 0.001\)); both within-group but outside-community and out-group structures showed negative associations. Taken together, these results indicate that recommendation feedback did not unfold uniformly, but rather propagated along community boundaries. In other words, structural bias in the recommendation system was not merely a static form of network differentiation, but a structural force that continued to shape subsequent pathways of information exposure.As a supplementary check, we further estimated the three-level lagged models separately for male and female nodes (see Appendix Fig.~\ref{fig:appendix_4}). We found that the overall dynamic pattern remains similar across groups, although the relative strength of within-community effects is more pronounced among female nodes in the click \(\rightarrow\) exposure stage.

\subsection{A Simple Model}

\begin{figure}[pos = htbp]
    \centering
    \begin{minipage}[b]{0.48\textwidth}
        \centering
        \includegraphics[width=\textwidth]{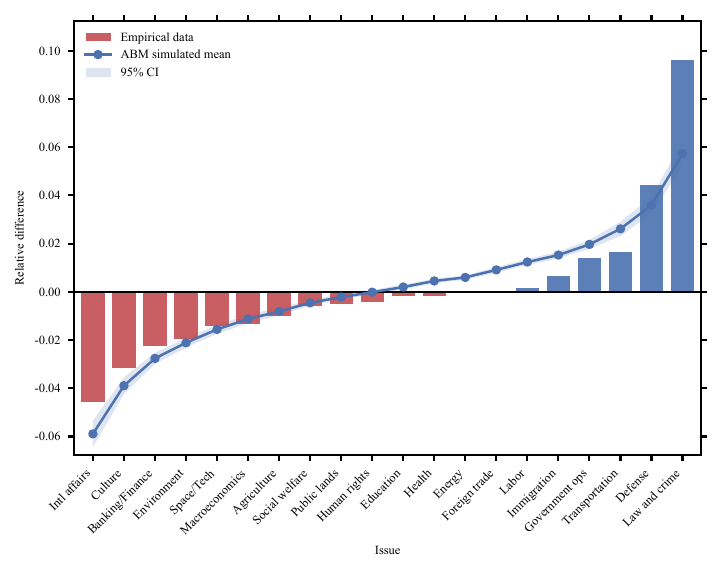}
    \end{minipage}
    \hfill
    \begin{minipage}[b]{0.48\textwidth}
        \centering
        \includegraphics[width=\textwidth]{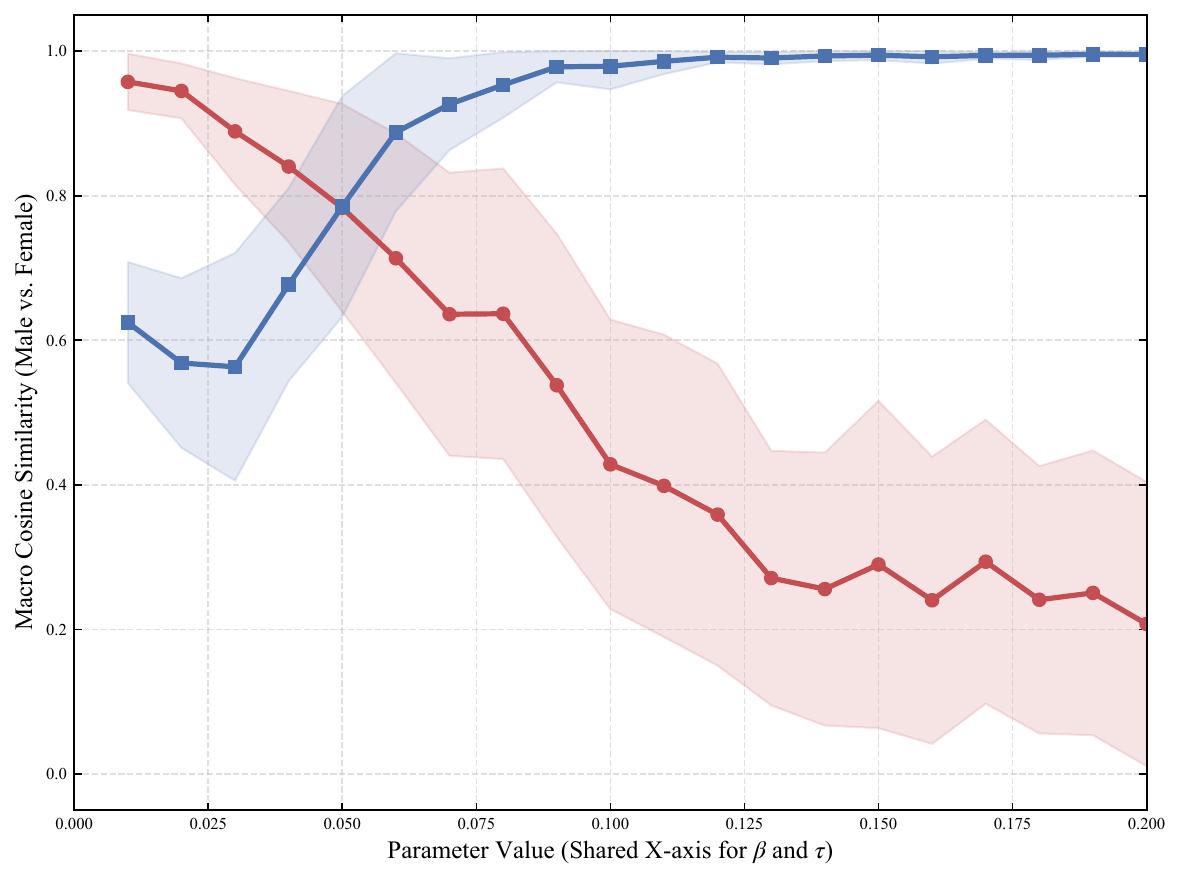}
    \end{minipage}
    \centerline{\makebox[0.45\textwidth][c]{(a)} \hspace{0.05\textwidth} \makebox[0.45\textwidth][c]{(b)}}
    \caption{Empirical--simulation comparison and parameter sensitivity analysis. Panel (a) compares the empirical gender differences in issue distribution with the mean results generated by the agent-based model (ABM). Gray bars represent empirical differences, the blue line indicates the mean simulated differences, and the shaded band denotes the 95\% confidence interval across simulation runs. Panel (b) presents the sensitivity analysis of the model parameters. It shows how group-level differentiation varies with changes in the gender-homophily parameter and the temperature parameter, illustrating how stronger homophily and lower recommendation randomness contribute to more pronounced divergence between gender-coded profiles.}
    \label{fig:figure6}
\end{figure}

As a complement to the empirical analysis, Study 2 used a simple model to examine, from a collaborative-filtering perspective, whether clear group differentiation could still gradually emerge over 150 time steps even in the absence of pre-specified initial differences in political preferences, thereby reproducing the issue differentiation observed on YouTube (Fig.~\ref{fig:figure6}). Building on the model's ability to reproduce the empirical pattern of differentiation, we further conducted a sensitivity analysis to clarify the mechanism associated with its core parameters. The results show that the gender-homophily parameter captures the extent to which the system assigns additional weight to accounts sharing the same gender-coded behavioral profile when calculating similarity. As this parameter increases, the similarity between the two groups in their macro-level issue centers declines and then stabilizes beyond a certain range. By contrast, the temperature parameter \(\tau\) does not represent collaborative filtering itself; rather, it regulates how concentrated similarity scores are in the recommendation sampling process. A lower value of \(\tau\) means that the system is more likely to distribute content from the most similar sources, thereby making group differentiation easier to reinforce. A higher value of \(\tau\), in contrast, preserves more exploration and randomness in recommendation, thus mitigating information separation between groups to some extent. Taken together, these results suggest that even when users enter the system without substantial initial differences, a slight preference for same-type profiles within a collaborative-filtering framework, combined with a low-exploration sampling mechanism, can gradually accumulate through repeated feedback and produce differentiated information environments. It should be noted that this model does not constitute a direct identification of YouTube's underlying mechanism; rather, it serves as a supplementary mechanistic demonstration showing that the empirical findings are theoretically generable.

\FloatBarrier

\section{Discussion}

Building on prior scholarship on gender bias in recommendation systems \citep{RN19,RN40}, political information exposure \citep{RN17}, and bias in platform distribution \citep{RN34}, this study used a controlled social-bot field experiment on YouTube to examine whether recommendation systems organize differentiated political information environments around gendered behavioral cues. Our findings suggest that recommendation systems not only shape the allocation of political content, but may also structure political information environments in uneven ways. Specifically, the political content recommended to male-coded and female-coded profiles revealed risks of allocative gender bias across issues, ideology, and entities, as well as structural gender bias in the organization of recommendation environments. Given the limited transparency of YouTube's model training and development processes, we further explored collaborative filtering as one possible mechanism underlying these gendered differences. Overall, the results underscore the need for sustained scrutiny and monitoring of gender bias in recommendation algorithms.

First, our findings provide evidence for the allocative dimension of gender bias in recommendation systems \citep{RN70}. Although YouTube produced broadly comparable levels of overall political exposure across gender-coded profiles, the internal composition of that exposure diverged substantially. This pattern is consistent with prior research suggesting that platform recommendations do not distribute content neutrally and that different groups may be directed toward different forms of exposure and visibility \citep{RN72,RN6}. The contribution of this study, however, is to show that allocative bias cannot be understood solely in terms of unequal exposure volume. Even when the intensity with which different profiles entered the political information stream was controlled as much as possible, the platform still generated different content allocations around gendered behavioral cues. In particular, even where differences in the proportion of political exposure were statistically significant but relatively limited in magnitude, the recommendation system continued to differentiate political content along issue distribution, ideological orientation, and entities. Importantly, this did not necessarily take the form of a straightforward overall disadvantage for female-coded profiles. In some cases, female-coded profiles were exposed to political content at rates comparable to, or even higher than, those of male-coded profiles, while still being allocated different issues and entities. This suggests that allocative bias in recommendation systems may be expressed less through bias in the sheer quantity of political exposure than through differentiated content allocation under broadly similar conditions of political contact. In this sense, our findings extend prior work on gender fairness and bias in recommendation systems \citep{RN19,RN61} by showing that gendered disparities may persist even when surface-level exposure differences remain limited. Accordingly, interpretations of platform gender bias based only on exposure share, visibility, or surface-level representation remain incomplete.

Second, whereas prior studies of gender bias in algorithms have largely concentrated on allocative bias, this study further examines structural bias in the organization of recommendation environments. Our results show that the political videos encountered by different gender-coded behavioral profiles were clearly differentiated in terms of within-group overlap, clustering, and community boundaries. This suggests that gender bias in recommendation systems is not only a matter of what content is allocated to whom, but also of how such content is connected, clustered, and organized \citep{RN76,RN66}. In this respect, our findings resonate with research on echo chambers, homophily, and clustered information environments, which has shown that platform-based differentiation does not always appear as straightforward ideological polarization, but may instead take the form of repeated circulation and reinforcement of similar content within established boundaries \citep{RN78,RN75,RN62}. Our analysis further indicates that these structural differences are not merely static. Rather, they continue to shape subsequent pathways of political video exposure through the feedback chain of exposure, click, and re-exposure, consistent with prior work on feedback loops and bias amplification in recommender systems \citep{RN77}. Once different profiles are directed into different community boundaries, subsequent recommendations are more likely to follow those existing structures, thereby reinforcing similar issues and narratives within groups while reducing opportunities for cross-group exposure and informational bridging. Recommendation bias, therefore, should be understood not only as a difference in content allocation, but also as a process through which platforms continuously organize pathways of political exposure around different profiles. Although this study does not directly infer specific political consequences from such structural bias, the findings suggest that platform recommendations may place different groups more stably into distinct political information environments through differentiated patterns of connection and feedback.

From a theoretical perspective, the contribution of this study lies not merely in identifying gender differences in both allocative bias and structural bias within YouTube's recommendation system, but in further showing that platform recommendations may be becoming an important organizing mechanism in the formation of identity-driven information ecologies \citep{RN70}. The core insight of the IDIE perspective is that different groups are not simply exposed to different kinds of information; rather, they are embedded in distinct information ecologies, and these ecological differences may, over time, translate into asymmetries in cognitive frameworks, public connection, and political opportunity \citep{RN62}. Our findings suggest that, in platformized recommendation environments, such bias is not merely a static reflection of pre-existing social differences, but may be continuously produced, organized, and reinforced through algorithmic filtering, curated flows, and iterative feedback \citep{RN77,RN42,RN76}. In other words, recommendation systems do not simply respond neutrally to identity differences. Instead, by simultaneously shaping content allocation and relational structure, they may continuously direct different gender-coded behavioral profiles into separate political information worlds, consistent with broader research on homophily, echo chambers, and identity-based differentiation in recommender systems \citep{RN78,RN73}. From this perspective, gender bias on platforms should not be understood only as unequal content distribution at the surface level, but as a form of bias at the level of information ecology: different groups are repeatedly placed into different issue spaces, connection structures, and conditions of public exposure, and through this process their asymmetric positions within political information environments are continually accumulated and reinforced \citep{RN66}.

\appendix
\section{Appendix}

\setcounter{figure}{0}
\renewcommand{\thefigure}{\arabic{figure}}
\renewcommand{\figurename}{Appendix Figure}

\setcounter{table}{0}
\renewcommand{\thetable}{\arabic{table}}
\renewcommand{\tablename}{Appendix Table}

\subsection{Prompt}

Here, we present the prompts that are used for querying the models.
Classify the following YouTube video titles as ideologically liberal, neutral, or conservative. Titles with no ideological content are classified as neutral. The news source is also specified for additional context. Only respond with the final answer \citep{RN24}.

Please tell me all the entity words in the text that belong to a given category. Output format is 'type1: word1; type2: word2' \citep{RN25}.

You are a topic classifier. Please carefully read the provided text and determine its topic based on the content.
If the article is most relevant to one of the topics in the list, return that topic.
If the article is not related to any topic in the list, return "Other."
Topic List:
Macroeconomics, Civil Rights, Health, Agriculture, Labor, Education, Environment, Energy, Immigration, Transportation, Law and Crime, Social Welfare, Housing, Domestic Commerce, Defense, Technology, Foreign Trade, International Affairs, Government Operations, Public Lands, Culture.
Please return the most relevant topic name or "Other". Store the result in the "issue" attribute.

\FloatBarrier
\subsection{Supplementary Figures}

\begin{figure}[pos = htbp]
    \centering
    \begin{minipage}[b]{0.25\textwidth}
        \centering
        \includegraphics[width=\textwidth]{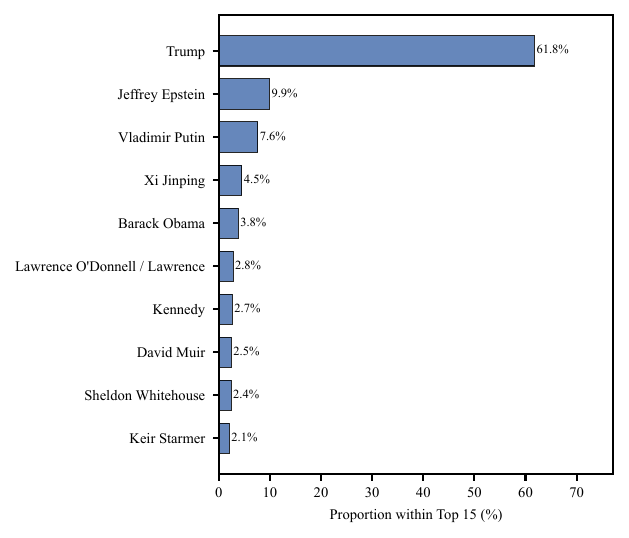}
        
        \small (a)
    \end{minipage}
    \begin{minipage}[b]{0.25\textwidth}
        \centering
        \includegraphics[width=\textwidth]{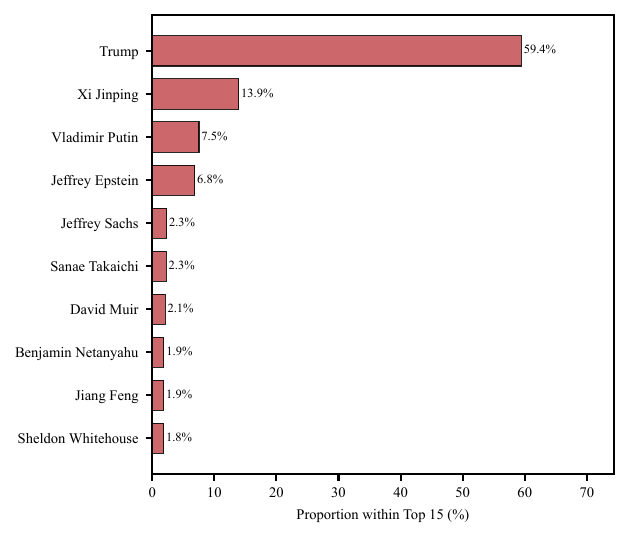}
        
        \small (b)
    \end{minipage}

    \vspace{0.6em}

    \begin{minipage}[b]{0.25\textwidth}
        \centering
        \includegraphics[width=\textwidth]{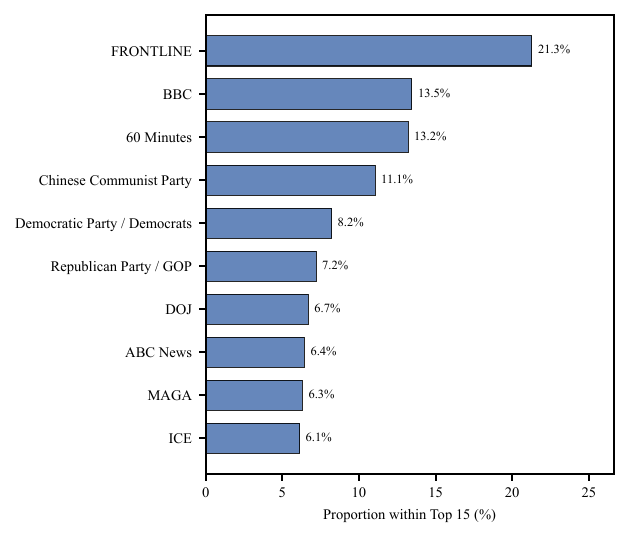}
        
        \small (c)
    \end{minipage}
    \begin{minipage}[b]{0.25\textwidth}
        \centering
        \includegraphics[width=\textwidth]{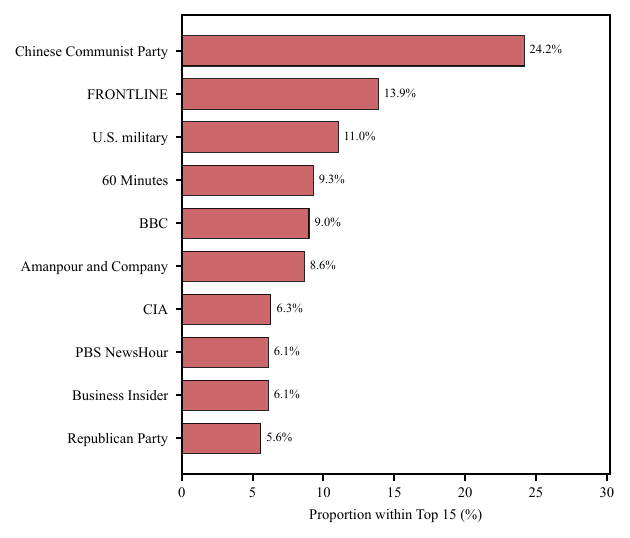}
        
        \small (d)
    \end{minipage}
    \caption{Top entities across gender-coded profiles. The figure reports the most frequently occurring person and organization entities in recommendation trajectories across male-coded and female-coded accounts. The upper panels present person entities, and the lower panels present organization entities. Values indicate the relative prominence of each entity within the corresponding group.}
    \label{fig:appendix1}
\end{figure}

\begin{figure}[pos = htbp]
    \centering
    \includegraphics[width=0.55\textwidth]{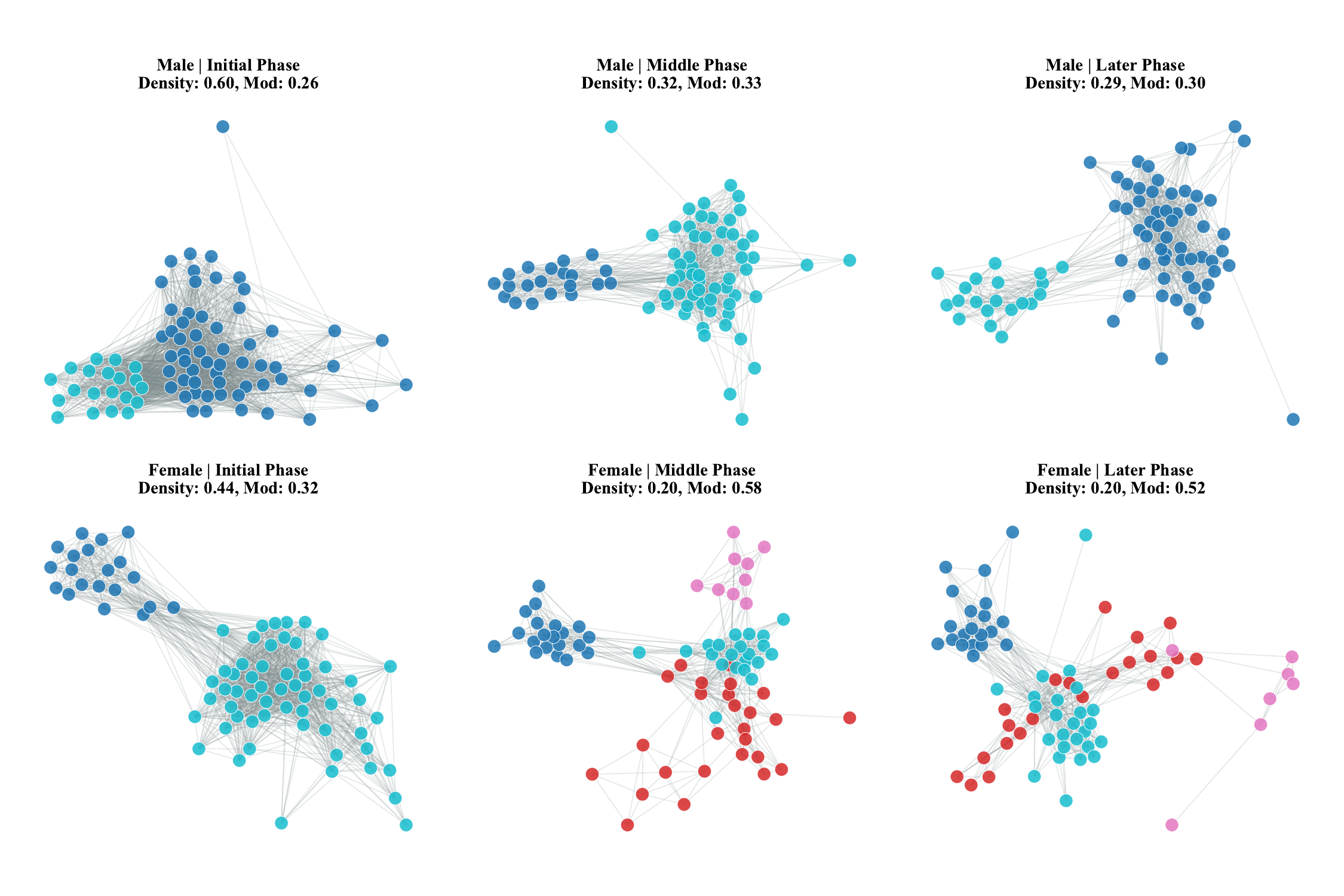}
    \caption{Changes in community structure across recommendation stages. The figure compares the community structure of co-exposure networks across different stages of the recommendation process, illustrating how modularity and community partitioning evolve from earlier to later interactions($\theta = 20$).}
    \label{fig:appendix2}
\end{figure}

\begin{figure}[pos = htbp]
    \centering
    \begin{minipage}[b]{0.36\textwidth}
        \centering
        \includegraphics[width=\textwidth]{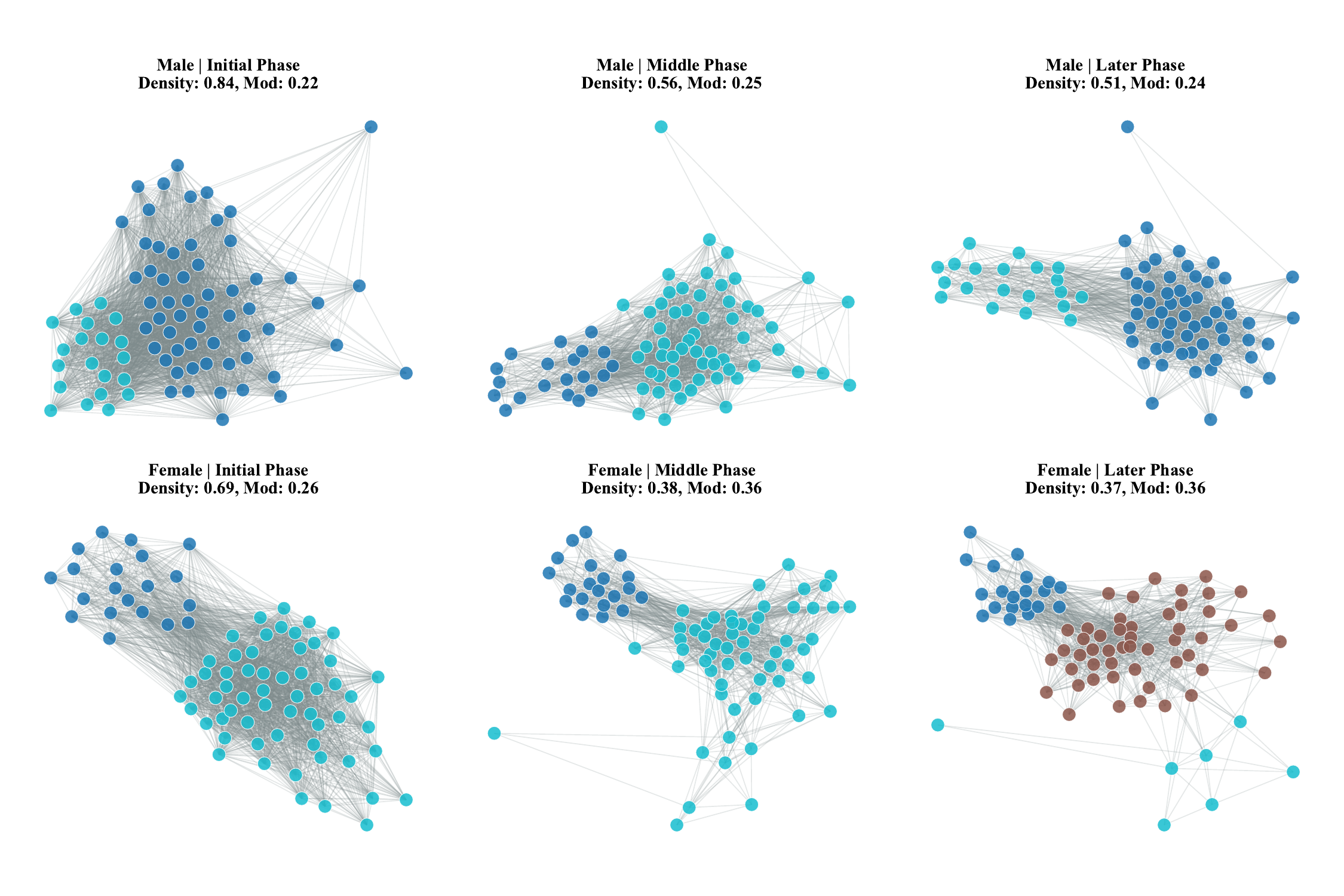}
        \label{fig:appendix2_10}
    \end{minipage}
    \begin{minipage}[b]{0.36\textwidth}
        \centering
        \includegraphics[width=\textwidth]{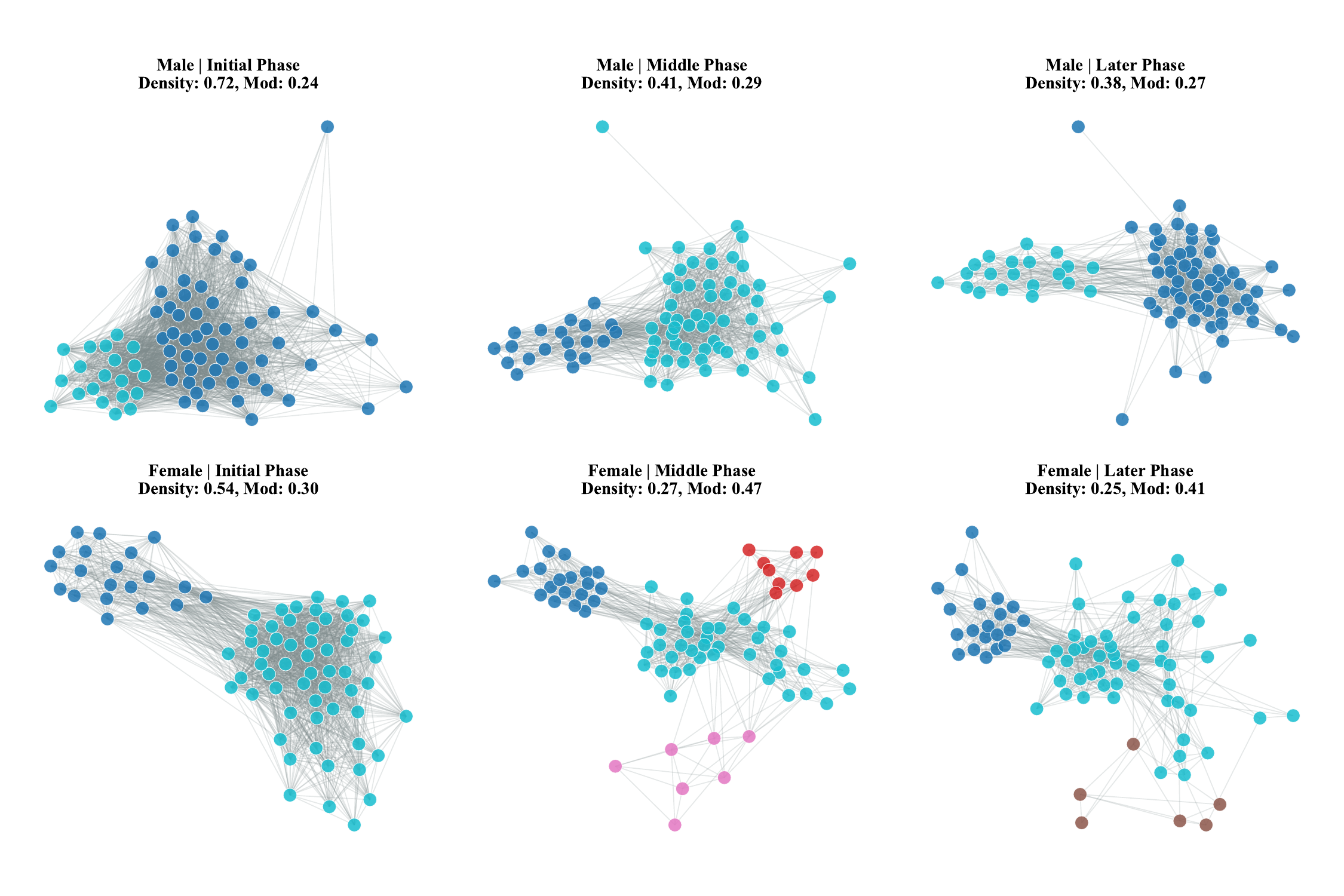}
        \label{fig:appendix2_15}
    \end{minipage}
    \centerline{\makebox[0.45\textwidth][c]{(a)} \hspace{0.05\textwidth} \makebox[0.45\textwidth][c]{(b)}}
    
    \vspace{0.6em} 
    \begin{minipage}[b]{0.36\textwidth}
        \centering
        \includegraphics[width=\textwidth]{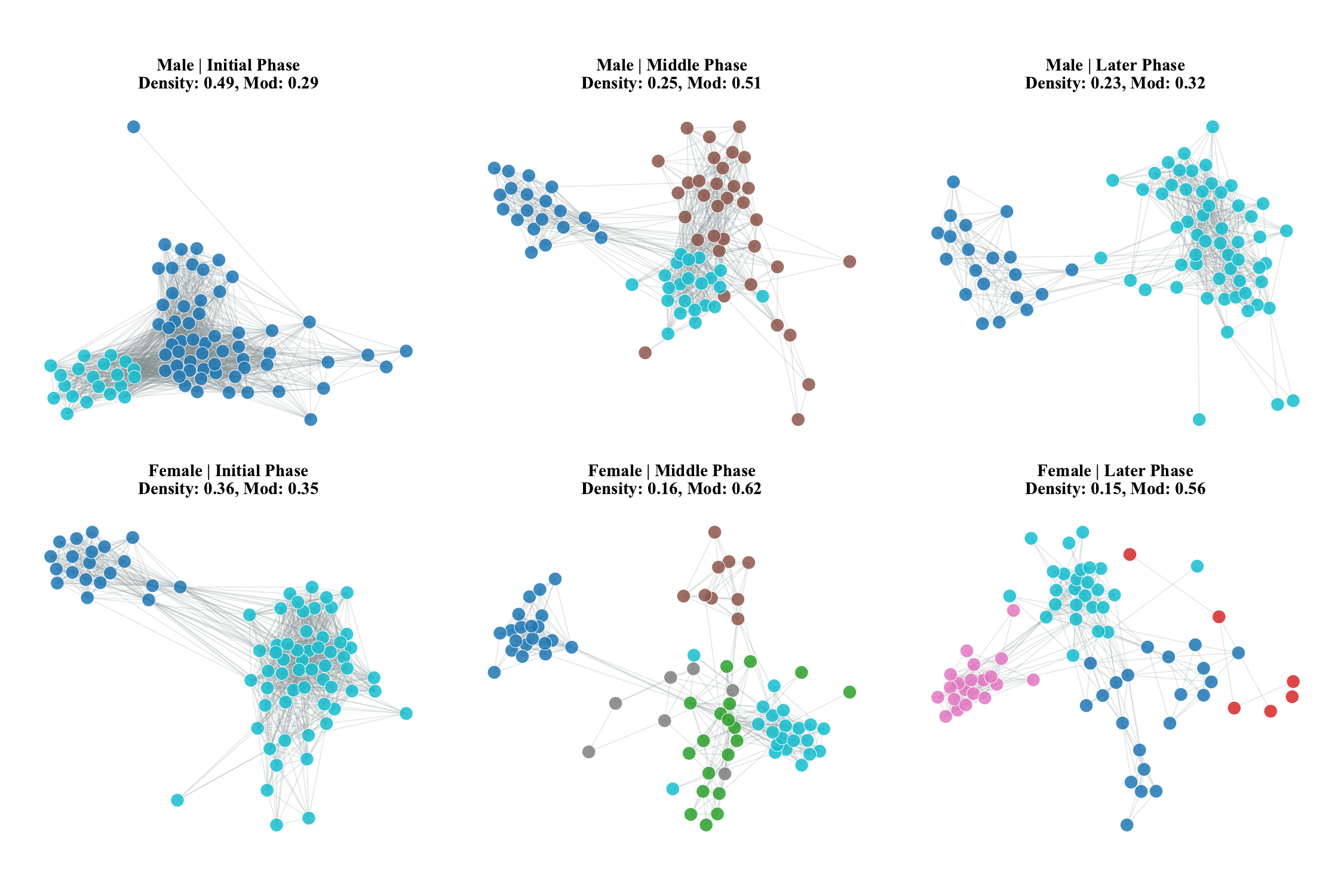}
        \label{fig:appendix2_20}
    \end{minipage}
    \begin{minipage}[b]{0.36\textwidth}
        \centering
        \includegraphics[width=\textwidth]{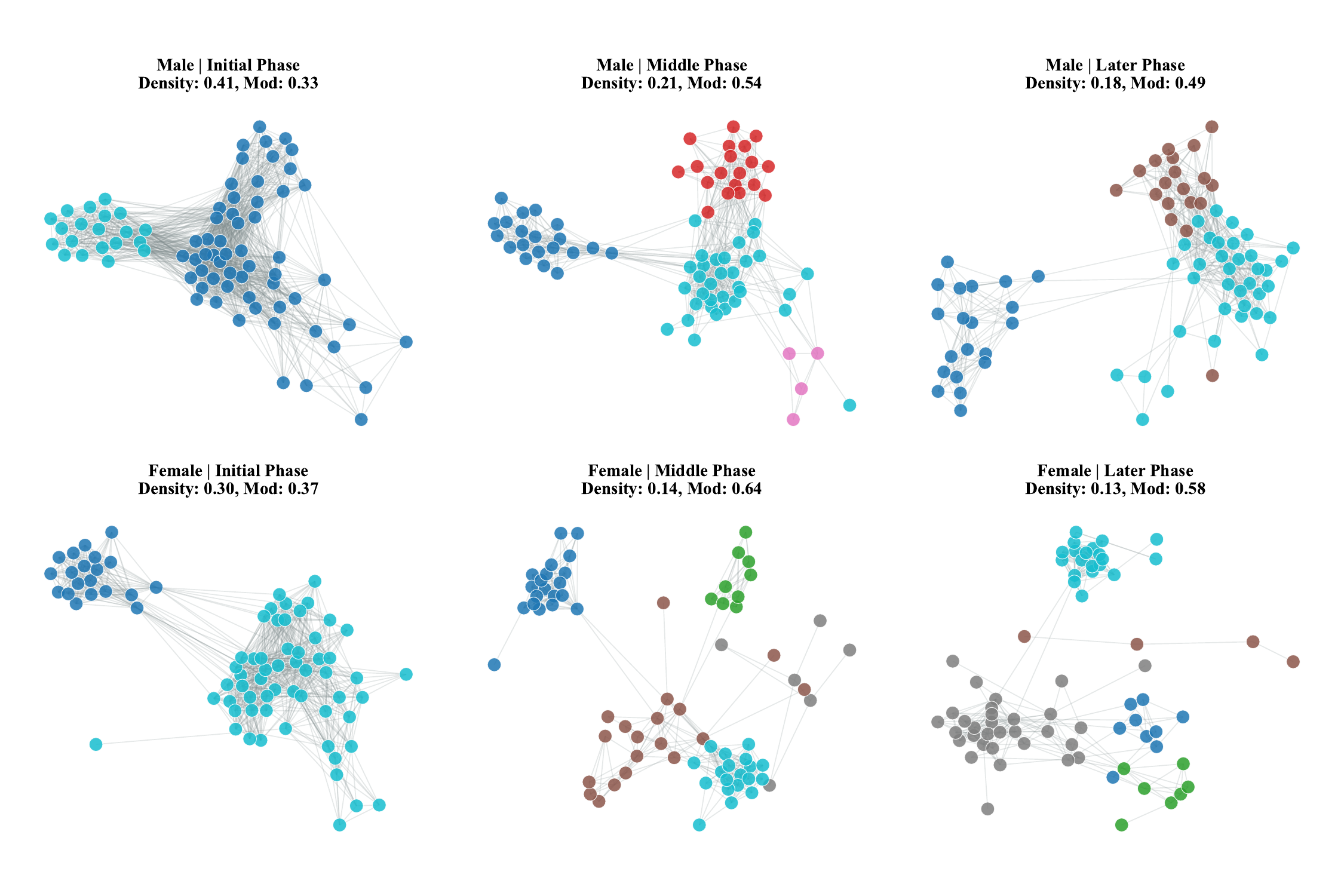}
        \label{fig:appendix2_25}
    \end{minipage}
    \centerline{\makebox[0.45\textwidth][c]{(c)} \hspace{0.05\textwidth} \makebox[0.45\textwidth][c]{(d)}}

    \caption{Changes in community structure across recommendation stages under different edge-weight thresholds ($\theta = 10, 15, 25, 30$). This figure compares the co-exposure networks across different stages of the recommendation process, illustrating how modularity and community partitioning evolve from earlier to later interactions.}
    \label{fig:appendix2_all}
\end{figure}

\begin{figure}[htbp]
    \centering

    \begin{minipage}[b]{0.48\textwidth}
        \centering
        \includegraphics[width=\textwidth]{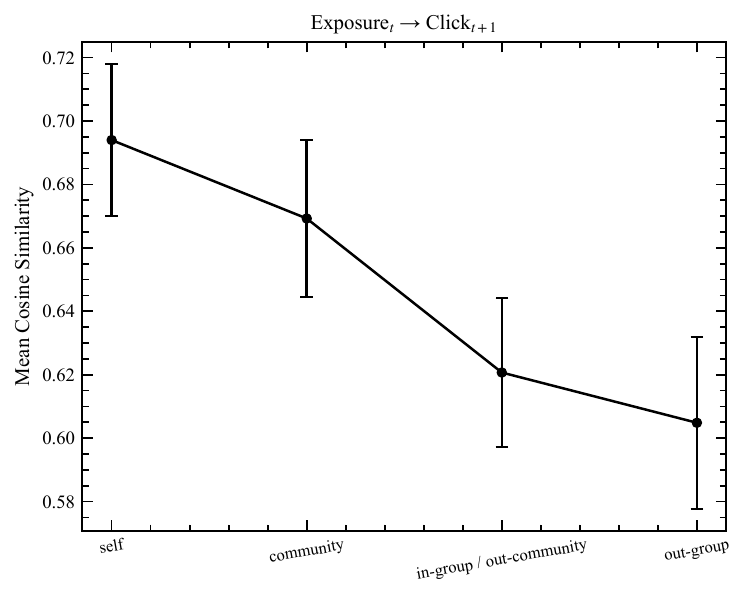}
        
        \small (a)
    \end{minipage}
    \hfill
    \begin{minipage}[b]{0.48\textwidth}
        \centering
        \includegraphics[width=\textwidth]{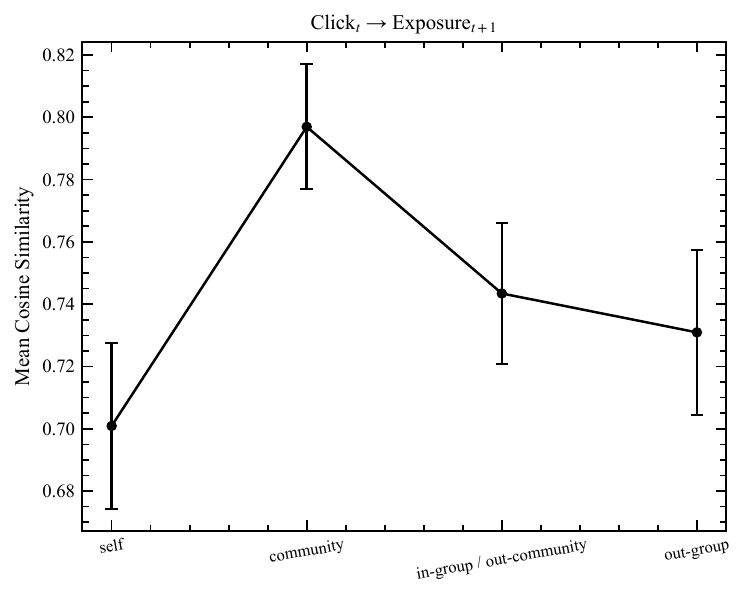}
        
        \small (b)
    \end{minipage}

    \vspace{0.6em}

    \begin{minipage}[b]{0.48\textwidth}
        \centering
        \includegraphics[width=\textwidth]{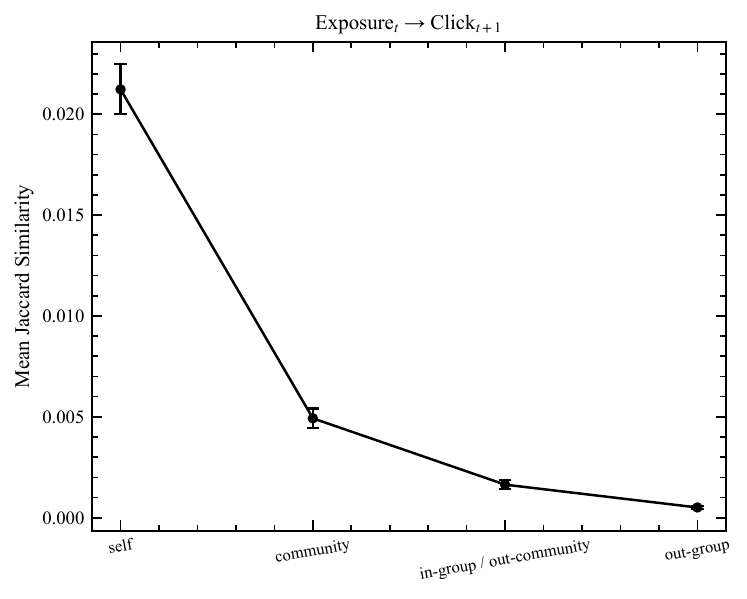}
        
        \small (c)
    \end{minipage}
    \hfill
    \begin{minipage}[b]{0.48\textwidth}
        \centering
        \includegraphics[width=\textwidth]{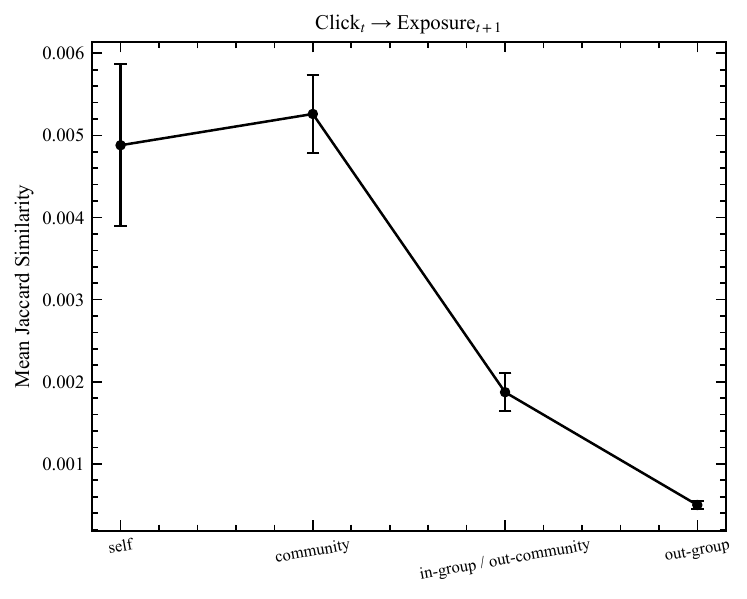}
        
        \small (d)
    \end{minipage}

    \caption{Similarity patterns across structural levels in the recommendation--click--re-recommendation process. Panels (a) and (b) report issue-level similarity, while panels (c) and (d) report ID-level similarity. Panels in the left column show the relationship between exposure structures at stage \(t\) and click outcomes at stage \(t+1\), whereas panels in the right column show the relationship between click structures at stage \(t\) and exposure outcomes at stage \(t+1\). Error bars indicate 95\% confidence intervals.}
    \label{fig:appendix3}
\end{figure}

\begin{figure}[h]
    \centering
    \begin{minipage}[b]{0.48\textwidth}
        \centering
        \includegraphics[width=\textwidth]{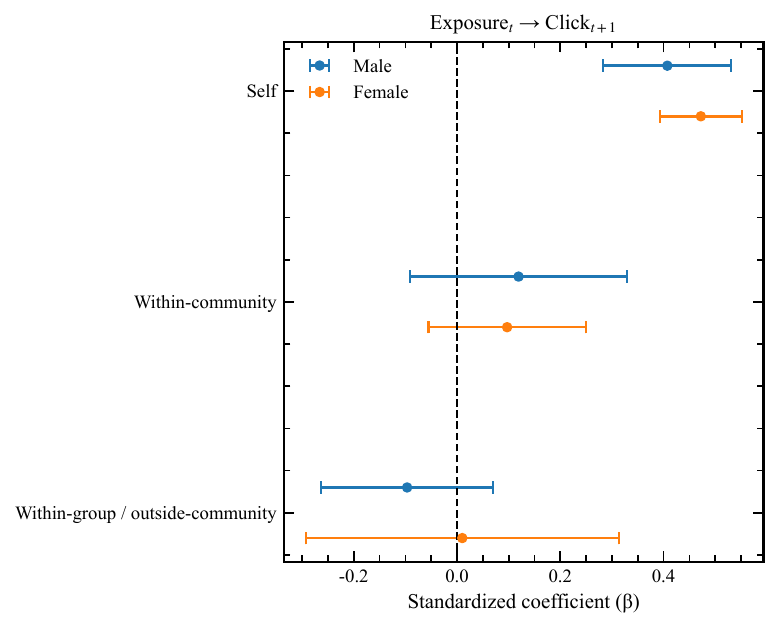}
    \end{minipage}
    \hfill
    \begin{minipage}[b]{0.48\textwidth}
        \centering
        \includegraphics[width=\textwidth]{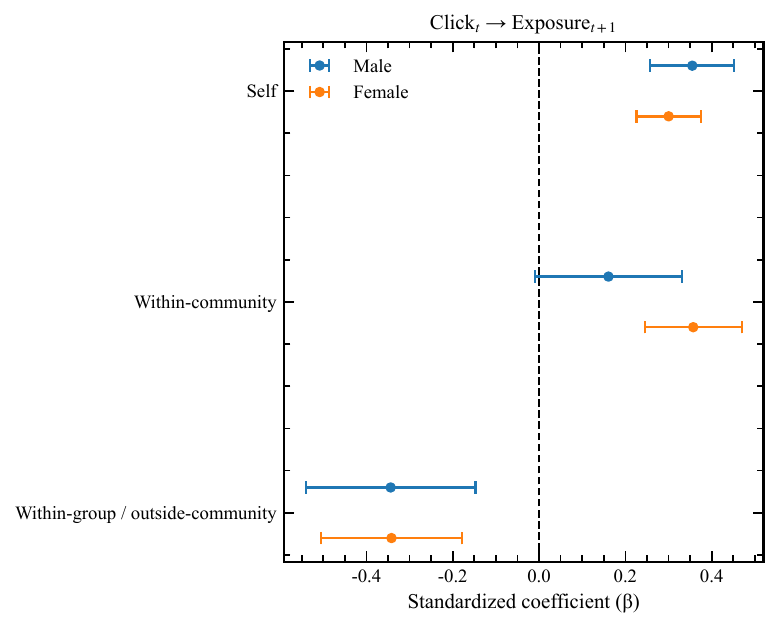}
    \end{minipage}
    \centerline{\makebox[0.45\textwidth][c]{(a)} \hspace{0.05\textwidth} \makebox[0.45\textwidth][c]{(b)}}

    \caption{Standardized coefficients from subgroup models of dynamic recommendation processes. Panel (a) reports subgroup estimates for Exposure$_t$ $\rightarrow$ Click$_{t+1}$ dynamics, and Panel (b) reports subgroup estimates for Click$_t$ $\rightarrow$ Exposure$_{t+1}$ dynamics. In both panels, coefficients are obtained from three-level models including self, within-community, and within-group/outside-community structures. All models include issue and stage fixed effects, with standard errors clustered at the node level. Points represent standardized coefficients ($\beta$), and horizontal bars indicate 95\% confidence intervals.}
    \label{fig:appendix_4}
\end{figure}

\FloatBarrier
\subsection{Supplementary Table}
\begin{table}[pos = h]
\caption{Gender Differences in the Exposure Ratios of \textit{News \& Politics} and \textit{Interest} Across Window Sizes}
\label{tab:appendix1}
\begin{tabular*}{\tblwidth}{@{} LLLLL @{}}
\toprule
Window & Category & Male (M$\pm$SD) & Female (M$\pm$SD) & P-value \\
\midrule
30 & News \& Politics & 0.133 $\pm$ 0.081 & 0.170 $\pm$ 0.106 & 0.0162* \\
30 & Interest         & 0.532 $\pm$ 0.134 & 0.485 $\pm$ 0.098 & 0.0147* \\
40 & News \& Politics & 0.134 $\pm$ 0.079 & 0.172 $\pm$ 0.102 & 0.0110* \\
40 & Interest         & 0.530 $\pm$ 0.132 & 0.483 $\pm$ 0.095 & 0.0120* \\
50 & News \& Politics & 0.134 $\pm$ 0.078 & 0.172 $\pm$ 0.101 & 0.00920** \\
50 & Interest         & 0.530 $\pm$ 0.132 & 0.481 $\pm$ 0.093 & 0.00955** \\
60 & News \& Politics & 0.134 $\pm$ 0.077 & 0.171 $\pm$ 0.098 & 0.00901** \\
60 & Interest         & 0.528 $\pm$ 0.133 & 0.480 $\pm$ 0.093 & 0.0109* \\
70 & News \& Politics & 0.134 $\pm$ 0.076 & 0.174 $\pm$ 0.096 & 0.00541** \\
70 & Interest         & 0.524 $\pm$ 0.134 & 0.478 $\pm$ 0.092 & 0.0148* \\
\bottomrule
\end{tabular*}
\vspace{2pt}
\begin{flushleft}
\footnotesize Note. M = mean; SD = standard deviation; Values are reported as M ± SD. P-values are based on two-tailed independent-samples t-tests. *\(p < 0.05\), **\(p < 0.01\), ***\(p < 0.001\).
\end{flushleft}
\end{table}

\begin{table}[pos = h]
\caption{Comparison of Exposure Diversity (Entropy Values) Between Male and Female Groups Across Different Window Sizes}
\label{tab:appendix2}
\footnotesize
\begin{tabular*}{\tblwidth}{@{\extracolsep{\fill}}lllll@{}}
\toprule
Window & Metric & Male (M$\pm$SD) & Female (M$\pm$SD) & P-value \\
\midrule
30 & Total Entropy  & 2.776 $\pm$ 0.364 & 2.651 $\pm$ 0.281 & 0.0187* \\
30 & Struct Entropy & 1.314 $\pm$ 0.212 & 1.391 $\pm$ 0.157 & 0.0114* \\
40 & Total Entropy  & 2.789 $\pm$ 0.366 & 2.664 $\pm$ 0.273 & 0.0177* \\
40 & Struct Entropy & 1.320 $\pm$ 0.205 & 1.400 $\pm$ 0.152 & 0.00686** \\
50 & Total Entropy  & 2.790 $\pm$ 0.370 & 2.677 $\pm$ 0.265 & 0.0300* \\
50 & Struct Entropy & 1.320 $\pm$ 0.207 & 1.403 $\pm$ 0.145 & 0.00433** \\
60 & Total Entropy  & 2.796 $\pm$ 0.376 & 2.684 $\pm$ 0.258 & 0.0326* \\
60 & Struct Entropy & 1.321 $\pm$ 0.207 & 1.405 $\pm$ 0.141 & 0.00376** \\
70 & Total Entropy  & 2.811 $\pm$ 0.361 & 2.694 $\pm$ 0.254 & 0.0213* \\
70 & Struct Entropy & 1.324 $\pm$ 0.199 & 1.411 $\pm$ 0.139 & 0.00223** \\
\bottomrule
\end{tabular*}
\vspace{2pt}
\begin{flushleft}
\footnotesize Note. Note. M = mean; SD = standard deviation; Values are reported as M ± SD. P-values are based on two-tailed independent-samples t-tests. *\(p < 0.05\), **\(p < 0.01\), ***\(p < 0.001\).
\end{flushleft}
\end{table}

\begin{table}[t]
\caption{Statistical Analysis of Political Issue Distribution Between Male and Female Groups}
\label{tab:appendix3}
\footnotesize
\begin{tabular*}{\tblwidth}{@{\extracolsep{\fill}}p{4.0cm}llll@{}}
\toprule
Issue & Male (M$\pm$SD) & Female (M$\pm$SD) & Diff (M$-$F) & P-value \\
\midrule
Law and crime              & 0.178 $\pm$ 0.131 & 0.094 $\pm$ 0.080 & 0.0833  & 0.000005*** \\
Defense                    & 0.245 $\pm$ 0.189 & 0.173 $\pm$ 0.135 & 0.0720  & 0.042538* \\
Transportation             & 0.045 $\pm$ 0.041 & 0.027 $\pm$ 0.041 & 0.0176  & 0.020256* \\
Government operations      & 0.102 $\pm$ 0.073 & 0.093 $\pm$ 0.065 & 0.0091  & 0.346704 \\
Immigration                & 0.017 $\pm$ 0.031 & 0.010 $\pm$ 0.018 & 0.0069  & 0.144700 \\
Energy                     & 0.023 $\pm$ 0.026 & 0.023 $\pm$ 0.033 & 0.0005  & 0.884692 \\
Community development      & 0.011 $\pm$ 0.016 & 0.013 $\pm$ 0.021 & -0.0017 & 0.596048 \\
Foreign trade              & 0.016 $\pm$ 0.027 & 0.018 $\pm$ 0.026 & -0.0018 & 0.939778 \\
Education                  & 0.006 $\pm$ 0.014 & 0.007 $\pm$ 0.016 & -0.0018 & 0.389892 \\
Labor and employment       & 0.011 $\pm$ 0.017 & 0.013 $\pm$ 0.020 & -0.0025 & 0.756402 \\
Health                     & 0.018 $\pm$ 0.037 & 0.022 $\pm$ 0.036 & -0.0034 & 0.693004 \\
Human rights               & 0.014 $\pm$ 0.020 & 0.019 $\pm$ 0.023 & -0.0044 & 0.2896 \\
Public lands \& water      & 0.003 $\pm$ 0.011 & 0.008 $\pm$ 0.026 & -0.0050 & 0.128343 \\
Social welfare             & 0.005 $\pm$ 0.015 & 0.012 $\pm$ 0.033 & -0.0073 & 0.120254 \\
Agriculture                & 0.007 $\pm$ 0.018 & 0.017 $\pm$ 0.026 & -0.0094 & 0.006990** \\
Macroeconomics             & 0.046 $\pm$ 0.049 & 0.060 $\pm$ 0.052 & -0.0147 & 0.076510 \\
Space, sci, tech \& comm   & 0.030 $\pm$ 0.036 & 0.047 $\pm$ 0.045 & -0.0165 & 0.022777* \\
Environment                & 0.010 $\pm$ 0.017 & 0.029 $\pm$ 0.050 & -0.0182 & 0.002260** \\
Culture and arts           & 0.049 $\pm$ 0.045 & 0.081 $\pm$ 0.082 & -0.0321 & 0.002732** \\
Banking and finance        & 0.050 $\pm$ 0.064 & 0.084 $\pm$ 0.113 & -0.0336 & 0.056172 \\
International affairs      & 0.114 $\pm$ 0.102 & 0.151 $\pm$ 0.102 & -0.0369 & 0.008734** \\
\bottomrule
\end{tabular*}
\vspace{2pt}
\begin{flushleft}
\footnotesize Note. M = mean; SD = standard deviation; Values are reported as M ± SD. P-values are based on two-tailed independent-samples t-tests. *\(p < 0.05\), **\(p < 0.01\), ***\(p < 0.001\).
\end{flushleft}
\end{table}

\begin{table}[pos = t]
\caption{Overall Structural Results}
\label{tab:appendix4}
\footnotesize
\begin{tabular*}{\tblwidth}{@{\extracolsep{\fill}}p{3.0cm}p{5.3cm}lll@{}}
\toprule
Model & Comparison & \(n\) & Mean difference & Holm-adjusted \(p\) \\
\midrule
Exposure \(\rightarrow\) Click   & self \(>\) community                         & 148 & 0.0290 & 0.0032** \\
Exposure \(\rightarrow\) Click   & community \(>\) in-group / out-community    & 148 & 0.0479 & \(< .001\)*** \\
Exposure \(\rightarrow\) Click   & community \(>\) out-group                    & 148 & 0.0700 & \(< .001\)*** \\
Exposure \(\rightarrow\) Click   & in-group / out-community \(>\) out-group     & 148 & 0.0222 & 0.0220* \\
Click \(\rightarrow\) Exposure   & community \(>\) self                         & 150 & 0.0933 & \(< .001\)*** \\
Click \(\rightarrow\) Exposure   & community \(>\) in-group / out-community    & 150 & 0.0524 & \(< .001\)*** \\
Click \(\rightarrow\) Exposure   & community \(>\) out-group                    & 150 & 0.0715 & \(< .001\)*** \\
Click \(\rightarrow\) Exposure   & in-group / out-community \(>\) out-group     & 150 & 0.0191 & 0.0172* \\
\bottomrule
\end{tabular*}
\end{table}

\begin{table}[t]
\caption{Hierarchical Feedback from Exposure to Subsequent Clicks (Issue Level)}
\label{tab:appendix5}
\footnotesize
\begin{tabular*}{\tblwidth}{@{\extracolsep{\fill}}llllll@{}}
\toprule
Group & Transition & Comparison & \(n\) & Mean difference & Holm-adjusted \(p\) \\
\midrule
Female & Part 1\(\rightarrow\)2 & self \(>\) community                      & 73 & 0.0158 & 0.1186 \\
Female & Part 1\(\rightarrow\)2 & community \(>\) in-group / out-community & 73 & 0.0190 & 0.1186 \\
Female & Part 1\(\rightarrow\)2 & community \(>\) out-group                 & 73 & 0.0699 & \(< .001\)*** \\
Female & Part 1\(\rightarrow\)2 & in-group / out-community \(>\) out-group  & 73 & 0.0509 & \(< .001\)*** \\
Female & Part 2\(\rightarrow\)3 & self \(>\) community                      & 73 & 0.0623 & 0.0036** \\
Female & Part 2\(\rightarrow\)3 & community \(>\) in-group / out-community & 73 & 0.0347 & 0.0071** \\
Female & Part 2\(\rightarrow\)3 & community \(>\) out-group                 & 73 & 0.0794 & \(< .001\)*** \\
Female & Part 2\(\rightarrow\)3 & in-group / out-community \(>\) out-group  & 73 & 0.0447 & 0.0036** \\
Male   & Part 1\(\rightarrow\)2 & self \(>\) community                      & 75 & -0.0007 & 0.6965 \\
Male   & Part 1\(\rightarrow\)2 & community \(>\) in-group / out-community & 75 & 0.0499 & 0.0012** \\
Male   & Part 1\(\rightarrow\)2 & community \(>\) out-group                 & 75 & 0.0574 & \(< .001\)*** \\
Male   & Part 1\(\rightarrow\)2 & in-group / out-community \(>\) out-group  & 75 & 0.0074 & 0.6965 \\
Male   & Part 2\(\rightarrow\)3 & self \(>\) community                      & 73 & 0.0393 & 0.0287* \\
Male   & Part 2\(\rightarrow\)3 & community \(>\) in-group / out-community & 73 & 0.0846 & \(< .001\)*** \\
Male   & Part 2\(\rightarrow\)3 & community \(>\) out-group                 & 73 & 0.0736 & \(< .001\)*** \\
Male   & Part 2\(\rightarrow\)3 & in-group / out-community \(>\) out-group  & 73 & -0.0110 & 0.6831 \\
\bottomrule
\end{tabular*}
\end{table}

\begin{table}[pos = t]
\caption{Hierarchical Feedback from Clicks to Subsequent Exposure (Issue Level)}
\label{tab:appendix6}
\footnotesize
\begin{tabular*}{\tblwidth}{@{\extracolsep{\fill}}llllll@{}}
\toprule
Group & Transition & Comparison & \(n\) & Mean difference & Holm-adjusted \(p\) \\
\midrule
Female & Part 1\(\rightarrow\)2 & community \(>\) self                      & 73 & 0.1004 & \(< .001\)*** \\
Female & Part 1\(\rightarrow\)2 & community \(>\) in-group / out-community & 73 & 0.0471 & 0.0012** \\
Female & Part 1\(\rightarrow\)2 & community \(>\) out-group                 & 73 & 0.0629 & \(< .001\)*** \\
Female & Part 1\(\rightarrow\)2 & in-group / out-community \(>\) out-group  & 73 & 0.0158 & 0.0213* \\
Female & Part 2\(\rightarrow\)3 & community \(>\) self                      & 73 & 0.1085 & \(< .001\)*** \\
Female & Part 2\(\rightarrow\)3 & community \(>\) in-group / out-community & 73 & 0.0437 & \(< .001\)*** \\
Female & Part 2\(\rightarrow\)3 & community \(>\) out-group                 & 73 & 0.0781 & \(< .001\)*** \\
Female & Part 2\(\rightarrow\)3 & in-group / out-community \(>\) out-group  & 73 & 0.0344 & 0.0041** \\
Male   & Part 1\(\rightarrow\)2 & community \(>\) self                      & 77 & 0.0973 & \(< .001\)*** \\
Male   & Part 1\(\rightarrow\)2 & community \(>\) in-group / out-community & 77 & 0.0502 & \(< .001\)*** \\
Male   & Part 1\(\rightarrow\)2 & community \(>\) out-group                 & 77 & 0.0600 & \(< .001\)*** \\
Male   & Part 1\(\rightarrow\)2 & in-group / out-community \(>\) out-group  & 77 & 0.0098 & 0.2557 \\
Male   & Part 2\(\rightarrow\)3 & community \(>\) self                      & 75 & 0.0610 & 0.0097** \\
Male   & Part 2\(\rightarrow\)3 & community \(>\) in-group / out-community & 76 & 0.0669 & \(< .001\)*** \\
Male   & Part 2\(\rightarrow\)3 & community \(>\) out-group                 & 76 & 0.0859 & \(< .001\)*** \\
Male   & Part 2\(\rightarrow\)3 & in-group / out-community \(>\) out-group  & 76 & 0.0190 & 0.1469 \\
\bottomrule
\end{tabular*}
\end{table}
\FloatBarrier

\printcredits

\bibliographystyle{cas-model2-names}

\bibliography{cas-refs}

@inproceedings{RN69,
   author = {Blodgett, Su Lin and Barocas, Solon and Daumé III, Hal and Wallach, Hanna},
   title = {Language (Technology) is Power: A Critical Survey of “Bias” in NLP},
   booktitle = {Proceedings of the 58th Annual Meeting of the Association for Computational Linguistics},
   publisher = {Association for Computational Linguistics},
   pages = {5454-5476},
   year = {2020},
   type = {Conference Proceedings}
}

@book{RN2,
  editor    = {Gillespie, Tarleton and Boczkowski, Pablo J. and Foot, Kirsten A.},
  title     = {Media Technologies: Essays on Communication, Materiality, and Society},
  publisher = {The MIT Press},
  year      = {2014},
  address   = {Cambridge, MA},
  isbn      = {9780262525374}
}

@inproceedings{RN40,
  title={Unmasking gender bias in recommendation systems and enhancing category-aware fairness},
  author={Kheya, Tahsin Alamgir and Bouadjenek, Mohamed Reda and Aryal, Sunil},
  booktitle={Proceedings of the ACM on Web Conference 2025},
  pages={5127--5138},
  year={2025}
}

@inproceedings{RN70,
   author = {Barocas, Solon and Crawford, Kate and Shapiro, Aaron and Wallach, Hanna},
   title = {The Problem with Bias: Allocative Versus Representational Harms in Machine Learning},
   booktitle = {9th Annual Conference of the Special Interest Group for Computing, Information and Society (SIGCIS)},
   address = {Philadelphia, PA},
   year = {2017},
   type = {Conference Proceedings}
}

@article{RN79,
   author = {Mehrabi, Ninareh and Morstatter, Fred and Saxena, Nripsuta and Lerman, Kristina and Galstyan, Aram},
   title = {A Survey on Bias and Fairness in Machine Learning},
   journal = {ACM Computing Surveys},
   volume = {54},
   number = {6},
   pages = {1-35},
   year = {2021},
   type = {Journal Article}
}

@article{RN74,
   author = {Prior, Markus},
   title = {News vs. Entertainment: How Increasing Media Choice Widens Gaps in Political Knowledge and Turnout},
   journal = {American Journal of Political Science},
   volume = {49},
   number = {3},
   pages = {577-592},
   ISSN = {0092-5853},
   year = {2005},
   type = {Journal Article}
}

@article{RN65,
   author = {Mitova, Eliza and Blassnig, Sina and Strikovic, Edina and Urman, Aleksandra and Hannak, Aniko and de Vreese, Claes H. and Esser, Frank},
   title = {News Recommender Systems: A Programmatic Research Review},
   journal = {Annals of the International Communication Association},
   volume = {47},
   number = {1},
   pages = {84-113},
   ISSN = {2380-8985},
   year = {2022},
   type = {Journal Article}
}

@article{RN25,
   author = {Wang, Xiao and Zhou, Weikang and Zu, Can and Xia, Han and Chen, Tianze and Zhang, Yuansen and Zheng, Rui and Ye, Junjie and Zhang, Qi and Gui, Tao},
   title = {Instructuie: Multi-task instruction tuning for unified information extraction},
   journal = {arXiv preprint arXiv:2304.08085},
   year = {2023},
   type = {Journal Article}
}

@article{RN62,
   author = {Hiaeshutter-Rice, Dan and Madrigal, Guadalupe and Ploger, Gavin and Carr, Sydney and Carbone, Mia and Battocchio, Ava Francesca and Soroka, Stuart},
   title = {Identity Driven Information Ecosystems},
   journal = {Communication Theory},
   volume = {34},
   number = {2},
   pages = {82-91},
   ISSN = {1468-2885},
   year = {2024},
   type = {Journal Article}
}

@article{RN66,
   author = {Thorson, Kjerstin and Wells, Chris},
   title = {Curated Flows: A Framework for Mapping Media Exposure in the Digital Age},
   journal = {Communication Theory},
   volume = {26},
   number = {3},
   pages = {309-328},
   ISSN = {1050-3293},
   year = {2016},
   type = {Journal Article}
}

@article{RN9,
   author = {Kelm, Ole and Neumann, Tim and Behrendt, Maike and Brenneis, Markus and Gerl, Katharina and Marschall, Stefan and Meißner, Florian and Harmeling, Stefan and Vowe, Gerhard and Ziegele, Marc},
   title = {How algorithmically curated online environments influence users’ political polarization: Results from two experiments with panel data},
   journal = {Computers in Human Behavior Reports},
   volume = {12},
   pages = {100343},
   ISSN = {2451-9588},
   year = {2023},
   type = {Journal Article}
}

@article{RN3,
   author = {Diakopoulos, Nicholas},
   title = {Algorithmic Accountability},
   journal = {Digital Journalism},
   volume = {3},
   number = {3},
   pages = {398-415},
   ISSN = {2167-0811},
   year = {2015},
   type = {Journal Article}
}

@article{RN76,
   author = {Bozdag, E.},
   title = {Bias in algorithmic filtering and personalization},
   journal = {Ethics and Information Technology},
   volume = {15},
   number = {3},
   pages = {209-227},
   ISSN = {1388-1957},
   year = {2013},
   type = {Journal Article}
}

@article{RN80,
   author = {Olteanu, Alexandra and Castillo, Carlos and Diaz, Fernando and Kcman, Emre},
   title = {Social Data: Biases, Methodological Pitfalls, and Ethical Boundaries},
   journal = {Frontiers in Big Data},
   volume = {2},
   number = {2},
   pages = {456527},
   year = {2019},
   type = {Journal Article}
}

@article{RN60,
  author  = {Kubiak, Emeric and Efremova, Maria I. and Baron, Simon and Frasca, Keely J.},
  title   = {Gender equity in hiring: Examining the effectiveness of a personality-based algorithm},
  journal = {Frontiers in Psychology},
  year    = {2023},
  volume  = {14},
  pages   = {1219865}
}

@article{RN21,
   author = {Tian, Weiqi and Ge, Jingshen},
   title = {Decoding the apple paradox: a critical discourse analysis of gender, technology, and nationalism in China’s digital space},
   journal = {Humanities and Social Sciences Communications},
   volume = {11},
   number = {1},
   pages = {1269},
   ISSN = {2662-9992},
   year = {2024},
   type = {Journal Article}
}

@article{RN19,
   author = {Melchiorre, Alessandro B. and Rekabsaz, Navid and Parada-Cabaleiro, Emilia and Brandl, Stefan and Lesota, Oleg and Schedl, Markus},
   title = {Investigating gender fairness of recommendation algorithms in the music domain},
   journal = {Information Processing \& Management},
   volume = {58},
   number = {5},
   pages = {102666},
   ISSN = {0306-4573},
   year = {2021},
   type = {Journal Article}
}

@article{RN61,
  author  = {Yalcin, Emre and Bilge, Alper},
  title   = {Evaluating unfairness of popularity bias in recommender systems: A comprehensive user-centric analysis},
  journal = {Information Processing \& Management},
  year    = {2022},
  volume  = {59},
  number  = {6},
  pages   = {103100}
}

@article{RN42,
  author  = {Zhou, Meizi and Zhang, Jingjing and Adomavicius, Gediminas},
  title   = {Longitudinal Impact of Preference Biases on Recommender Systems' Performance},
  journal = {Information Systems Research},
  year    = {2024},
  volume  = {35},
  number  = {4},
  pages   = {1634--1656}
}

@article{RN6,
   author = {Helberger, Natali and Karppinen, Kari and D’Acunto, Lucia},
   title = {Exposure diversity as a design principle for recommender systems},
   journal = {Information, Communication \& Society},
   volume = {21},
   number = {2},
   pages = {191-207},
   ISSN = {1369-118X},
   year = {2018},
   type = {Journal Article}
}

@article{RN4,
   author = {Thelwall, Mike and Foster, David},
   title = {Male or female gender‐polarized YouTube videos are less viewed},
   journal = {Journal of the Association for Information Science and Technology},
   volume = {72},
   number = {12},
   pages = {1545–1557},
   ISSN = {2330-1635},
   year = {2021},
   type = {Journal Article}
}

@article{RN57,
   author = {Pethig, Florian and Kroenung, Julia},
   title = {Biased Humans, (Un)Biased Algorithms?},
   journal = {Journal of Business Ethics},
   volume = {183},
   number = {3},
   pages = {637-652},
   ISSN = {1573-0697},
   year = {2023},
   type = {Journal Article}
}

@article{RN63,
   author = {Rohrbach, Tobias and Aaldering, Loes and Van der Pas, Daphne Joanna},
   title = {Gender differences and similarities in news media effects on political candidate evaluations: a meta-analysis},
   journal = {Journal of Communication},
   volume = {73},
   number = {2},
   pages = {101-112},
   ISSN = {0021-9916},
   year = {2022},
   type = {Journal Article}
}

@article{RN59,
   author = {Van der Pas, Daphne Joanna and Aaldering, Loes},
   title = {Gender Differences in Political Media Coverage: A Meta-Analysis},
   journal = {Journal of Communication},
   volume = {70},
   number = {1},
   pages = {114-143},
   ISSN = {0021-9916},
   year = {2020},
   type = {Journal Article}
}

@article{RN64,
  author  = {Sun, Luhang and Wei, Mian and Sun, Yibing and Suh, Yoo Ji and Shen, Liwei and Yang, Sijia},
  title   = {Smiling women pitching down: Auditing representational and presentational gender biases in image-generative {AI}},
  journal = {Journal of Computer-Mediated Communication},
  year    = {2024},
  volume  = {29},
  number  = {1},
  ISSN = {1083-6101},
  pages   = {zmad045}
}

@article{RN58,
   author = {Rathee, S. and Banker, S. and Mishra, A. and Mishra, H.},
   title = {Algorithms propagate gender bias in the marketplace-with consumers' cooperation},
   journal = {Journal of Consumer Psychology},
   volume = {33},
   number = {4},
   pages = {621-631},
   ISSN = {1057-7408},
   year = {2023},
   type = {Journal Article}
}

@article{RN73,
   author = {Carrasco-Farré, Carlos and Grimaldi, Didier and Torrens, Marc and Longobuco, Enzo},
   title = {Social Identity Theory and Algorithmic Bias: Ingroup and Outgroup Acrophily in Recommender Systems},
   journal = {Journal of Management Information Systems},
   volume = {42},
   number = {4},
   pages = {1017-1054},
   ISSN = {0742-1222},
   year = {2025},
   type = {Journal Article}
}

@article{RN34,
   author = {Lambrecht, A. and Tucker, C.},
   title = {Algorithmic Bias? An Empirical Study of Apparent Gender-Based Discrimination in the Display of STEM Career Ads},
   journal = {Management Science},
   volume = {65},
   number = {7},
   pages = {2966-2981},
   ISSN = {0025-1909},
   year = {2019},
   type = {Journal Article}
}

@article{RN71,
   author = {Guilbeault, Douglas and Delecourt, Solène and Hull, Tasker and Desikan, Bhargav Srinivasa and Chu, Mark and Nadler, Ethan},
   title = {Online images amplify gender bias},
   journal = {Nature},
   volume = {626},
   number = {8001},
   pages = {1049-1055},
   ISSN = {1476-4687},
   year = {2024},
   type = {Journal Article}
}

@article{RN22,
   author = {Ulloa, Roberto and Richter, Ana Carolina and Makhortykh, Mykola and Urman, Aleksandra and Kacperski, Celina Sylwia},
   title = {Representativeness and face-ism: Gender bias in image search},
   journal = {New Media \& Society},
   volume = {26},
   number = {6},
   pages = {3541-3567},
   year = {2024},
   type = {Journal Article}
}

@article{RN45,
  author  = {Milli, Smitha and Carroll, Micah and Wang, Yike and Pandey, Sashrika and Zhao, Sebastian and Dragan, Anca D.},
  title   = {Engagement, user satisfaction, and the amplification of divisive content on social media},
  journal = {PNAS Nexus},
  year    = {2025},
  volume  = {4},
  number  = {3},
  pages   = {pgaf062}
}

@article{RN44,
  author  = {Yu, Xudong and Haroon, Muhammad and Menchen-Trevino, Ericka and Wojcieszak, Magdalena},
  title   = {Nudging recommendation algorithms increases news consumption and diversity on YouTube},
  journal = {PNAS Nexus},
  year    = {2024},
  volume  = {3},
  number  = {12},
  pages   = {pgae518}
}

@misc{RN24,
      title={"Whose Side Are You On?" Estimating Ideology of Political and News Content Using Large Language Models and Few-shot Demonstration Selection}, 
      author={Muhammad Haroon and Magdalena Wojcieszak and Anshuman Chhabra},
      year={2025},
      eprint={2503.20797},
      archivePrefix={arXiv},
      primaryClass={cs.CL},
}

@inproceedings{RN72,
  author    = {Singh, Ashudeep and Joachims, Thorsten},
  title     = {Fairness of Exposure in Rankings},
  booktitle = {Proceedings of the 24th {ACM} {SIGKDD} International Conference on Knowledge Discovery \& Data Mining},
  series    = {{KDD} '18},
  year      = {2018},
  pages     = {2219--2228},
  publisher = {Association for Computing Machinery},
  address   = {New York, NY, USA}
}

@inproceedings{RN41,
  author    = {Wang, Clarice and Wang, Kathryn and Bian, Andrew Y. and Islam, Rashidul and Keya, Kamrun Naher and Foulds, James R. and Pan, Shimei},
  title     = {Do Humans Prefer Debiased AI Algorithms? A Case Study in Career Recommendation},
  booktitle = {Proceedings of the 27th International Conference on Intelligent User Interfaces},
  series    = {IUI '22},
  year      = {2022},
  pages     = {134--147},
  address   = {Helsinki, Finland},
  publisher = {Association for Computing Machinery}
}

@inproceedings{RN77,
  author    = {Mansoury, Masoud and Abdollahpouri, Himan and Pechenizkiy, Mykola and Mobasher, Bamshad and Burke, Robin},
  title     = {Feedback Loop and Bias Amplification in Recommender Systems},
  booktitle = {Proceedings of the 29th {ACM} International Conference on Information \& Knowledge Management},
  series    = {{CIKM} '20},
  year      = {2020},
  pages     = {2145--2148},
  publisher = {Association for Computing Machinery},
  address   = {New York, NY, USA}
}

@inproceedings{RN11,
  author    = {Kay, Matthew and Matuszek, Cynthia and Munson, Sean A.},
  title     = {Unequal Representation and Gender Stereotypes in Image Search Results for Occupations},
  booktitle = {Proceedings of the 33rd Annual {ACM} Conference on Human Factors in Computing Systems},
  series    = {{CHI} '15},
  year      = {2015},
  pages     = {3819--3828},
  publisher = {Association for Computing Machinery},
  address   = {New York, NY, USA}
}

@article{RN17,
   author = {Huszár, Ferenc and Ktena, Sofia Ira and O’Brien, Conor and Belli, Luca and Schlaikjer, Andrew and Hardt, Moritz},
   title = {Algorithmic amplification of politics on Twitter},
   journal = {Proceedings of the National Academy of Sciences of the United States of America},
   volume = {119},
   number = {1},
   pages = {e2025334119},
   year = {2022},
   type = {Journal Article}
}

@article{RN8,
   author = {Vlasceanu, Madalina and Amodio, David M.},
   title = {Propagation of societal gender inequality by internet search algorithms},
   journal = {Proceedings of the National Academy of Sciences of the United States of America},
   volume = {119},
   number = {29},
   pages = {e2204529119},
   year = {2022},
   type = {Journal Article}
}

@article{RN78,
  author  = {Cinelli, Matteo and De Francisci Morales, Gianmarco and Galeazzi, Alessandro and Quattrociocchi, Walter and Starnini, Michele},
  title   = {The echo chamber effect on social media},
  journal = {Proceedings of the National Academy of Sciences of the United States of America},
  year    = {2021},
  volume  = {118},
  number  = {9},
  pages   = {e2023301118}
}

@inproceedings{RN75,
  author    = {Liu, Ping and Shivaram, Karthik and Culotta, Aron and Shapiro, Matthew A. and Bilgic, Mustafa},
  title     = {The Interaction between Political Typology and Filter Bubbles in News Recommendation Algorithms},
  booktitle = {Proceedings of the Web Conference 2021},
  series    = {{WWW} '21},
  year      = {2021},
  pages     = {3791--3801},
  publisher = {Association for Computing Machinery},
  address   = {New York, NY, USA}
}

@article{RN13,
   author = {Flaxman, Seth and Goel, Sharad and Rao, Justin M.},
   title = {Filter Bubbles, Echo Chambers, and Online News Consumption},
   journal = {Public Opinion Quarterly},
   volume = {80},
   number = {S1},
   pages = {298-320},
   ISSN = {0033-362X},
   year = {2016},
   type = {Journal Article}
}

@article{RN67,
   author = {Bakshy, Eytan and Messing, Solomon and Adamic, Lada A.},
   title = {Exposure to ideologically diverse news and opinion on Facebook},
   journal = {Science},
   volume = {348},
   number = {6239},
   pages = {1130-1132},
   year = {2015},
   type = {Journal Article}
}

@article{WOS:001439101300001,
  author  = {Liu, Naijia and Hu, Xinlan Emily and Savas, Yasemin and Baum, Matthew A. and Berinsky, Adam J. and Chaney, Allison J. B. and Lucas, Christopher and Mariman, Rei and de Benedictis-Kessner, Justin and Guess, Andrew M. and Knox, Dean and Stewart, Brandon M.},
  title   = {Short-term exposure to filter-bubble recommendation systems has limited polarization effects: Naturalistic experiments on {YouTube}},
  journal = {Proceedings of the National Academy of Sciences of the United States of America},
  year    = {2025},
  volume  = {122},
  number  = {8},
  pages   = {e2318127122}
}

@article{WOS:001128094700001,
    Author = {Haroon, Muhammad and Wojcieszak, Magdalena and Chhabra, Anshuman and
       Liu, Xin and Mohapatra, Prasant and Shafiq, Zubair},
    Title = {Auditing YouTube's recommendation system for ideologically congenial,
       extreme, and problematic recommendations},
    Journal = {Proceedings of the National Academy of Sciences of the United States of America},
    year    = {2023},
    volume  = {120},
    number  = {50},
    pages   = {e2213020120},
    Month = {DEC 5},
    ISSN = {0027-8424},
    EISSN = {1091-6490},
    ResearcherID-Numbers = {Wojcieszak, Magdalena/ABE-7590-2020},
}

@online{richmond2016openslate,
  author  = {Richmond, Will},
  title   = {OpenSlate Data Shows Vastly Different YouTube Viewing Interests for Young Men and Women},
  date    = {2016-05-08},
  organization = {VideoNuze},
  url     = {https://www.videonuze.com/article/openslate-data-shows-vastly-different-youtube-viewing-interests-for-young-men-and-women},
  urldate = {2026-04-28}
}

@online{cap_master_codebook,
  author  = {{Comparative Agendas Project}},
  title   = {Master Codebook},
  organization = {Comparative Agendas Project},
  url     = {https://www.comparativeagendas.net/pages/master-codebook},
  urldate = {2026-04-28}
}


\end{document}